\documentclass[10pt]{iopart}
\usepackage{iopams}  
\usepackage{cite}
\usepackage{epsfig}
\usepackage{graphicx}

\makeatletter
\def\NAT@parse{\typeout{This is a fake Natbib command to fool Hyperref.}}
\makeatother
\usepackage[hyperref]{hyperref}

\def\hc{{\cal H}_{\mathrm{C}}}
\def\hr{{\cal H}_{\mathrm{R}}}
\def\hleft{{\cal H}_{\ell}}
\def\hright{{\cal H}_{r}}
\def\hlr{{\cal H}_{\ell,r}}
\newcommand{\EEC}{\mathrm{EEC}}
\newcommand{\as}{\alpha_s}
\newcommand{\order}[1]{{\cal O}\left(#1\right)}
\newcommand{\ee}{e^+e^-}
\newcommand{\cP}{{\cal P}}
\newcommand{\cM}{{\cal M}}
\newcommand{\MSbar}{\overline{\mbox{\scriptsize MS}}}

\begin{document}

\begin{flushright}
 CERN-TH/2003-306\\
 LPTHE-03-40\\
 hep-ph/0312283
\end{flushright}

\title[Event shapes in $e^+e^-$ annihilation and deep inelastic scattering]{Event shapes in $\boldsymbol{e^+e^-}$ annihilation and deep inelastic scattering}

\author{Mrinal Dasgupta\dag\ and Gavin~P.~Salam\ddag}

\address{\dag\ CERN, Theory Division, CH-1211 Geneva 23, Switzerland.}

\address{\ddag\ LPTHE, Universities of Paris VI and VII, and CNRS UMR
  7589, Paris 75005, France.}

\begin{abstract}
  This article reviews the status of event-shape studies in $\ee$
  annihilation and
  DIS. It includes discussions of perturbative calculations, of
  various approaches to modelling hadronisation and of comparisons to
  data.
\end{abstract}



\section{Introduction}

\setcounter{footnote}{0}

Event shape variables are perhaps the most popular observables for
testing QCD and for improving our understanding of its dynamics.
Event shape studies began in earnest towards the late seventies as a
simple quantitative method to understand the nature of gluon
bremsstrahlung
\cite{EllisGaillardRoss,Farhi,GeorgMach,FoxWolf,Basham}. 
For instance, it was on the basis of comparisons to event shape data
that one could first deduce that gluons were vector particles, since
theoretical predictions employing scalar gluons did not agree with
experiment~\cite{ESWBook}.

Quite generally, event shapes parametrise the geometrical properties
of the energy-momentum flow of
an event and their values are therefore directly related to the
appearance of an event in the detector.  In other words the value of a
given event shape encodes in a continuous fashion, for example, the
transition from pencil-like two-jet events with hadron flow
prominently distributed along some axis, to planar three-jet events or
events with a spherical distribution of hadron momenta.  Thus they
provide more detailed information on the final state geometry than say
a jet finding algorithm which would always classify an event as having
a certain finite number of jets, even if the actual energy flow is
uniformly distributed in the detector and there is no prominent jet
structure present in the first place.

Event shapes are well suited to testing QCD mainly because, by
construction, they are collinear and infrared safe observables. This
means that one can safely compute them in perturbation theory and use
the predictions as a means of extracting the strong coupling
$\alpha_s$.  They are also well suited for determinations of other
parameters of the theory such as constraining the quark and gluon
colour factors as well as the QCD beta function.  Additionally they
have also been used in other studies for characterising the final
state, such as investigations of jet and heavy quark multiplicities as
functions of event shape variables~\cite{MikeDJM1,MikeDJM2}.

Aside from testing the basic properties of QCD, event shape
distributions are a powerful probe of our more detailed knowledge of
QCD dynamics.  All the commonly studied event shape variables have the
property that in the region where the event shape value is small, one
is sensitive primarily to gluon emission that is soft compared to the
hard scale of the event and/or collinear to one of the hard partons.
Such small-transverse-momentum emissions have relatively large
emission probabilities (compared to their high transverse momentum
counterparts) due to logarithmic soft-collinear dynamical enhancements
as well as the larger value of their coupling to the hard partons.
Predictions for event shape distributions therefore typically contain
large logarithms in the region where the event-shape is small, which
are a reflection of the importance of multiple soft and/or collinear
emission.  A successful prediction for an event shape in this region
requires all-order resummed perturbative predictions or the
use of a Monte-Carlo event generator, which contains correctly
the appropriate dynamics governing multiple particle production.
Comparisons of these predictions to experimental data are therefore a
stringent test of the understanding QCD dynamics that has been reached
so far.

One other feature of event shapes, that at first appeared as an
obstacle to their use in extracting the fundamental parameters of QCD,
is the presence of significant non-perturbative effects in the form of
power corrections that vary as an inverse power of the hard scale,
$(\Lambda/Q)^2p$.  For most event shapes, phenomenologically it was
found that $p=1/2$ (as discussed in~\cite{Webber94Tube}), and the
resulting non-perturbative effects can be of comparable size to
next-to-leading order perturbative predictions, as we shall discuss
presently.  The problem however, can be handled to a large extent by
hadronisation models embedded in Monte-Carlo event generators
\cite{Herwig,Pythia,Ariadne}, which model the conversion of partons
into hadrons at the cost of introducing several parameters that need
to be tuned to the data. Once this is done, application of such
hadronisation models leads to very successful comparisons of
perturbative predictions with experimental data with, for example,
values of $\alpha_s$ consistent to those obtained from other methods
and with relatively small errors.

However since the mid-nineties attempts have also been made to obtain
a better insight into the physics of hadronisation and power
corrections in particular.  Theoretical models such as those based on
renormalons have been developed to probe the non-perturbative domain
that gives rise to power corrections (for a review, see
\cite{Beneke:1998ui}).  Since the power corrections are relatively
large effects for event shape variables, scaling typically as $1/Q$
(no other class of observable shows such large effects), event shapes
have become the most widely used means of investigating the validity
of these ideas. In fact an entire phenomenology of event shapes has
developed which is based on accounting for non-perturbative effects
via such theoretically inspired models, and including them in fits to
event shape data alongside the extraction of other standard QCD
parameters. In this way event shapes also serve as a tool for
understanding more quantitatively the role of confinement effects.

The layout of this article is as follows. In the following section we
list the definitions of several commonly studied event shapes in
$e^{+}e^{-}$ annihilation and deep inelastic scattering (DIS) and
discuss briefly some of their
properties. Then in section~\ref{sec:perturbative} we review the state
of the art for perturbative predictions of event-shape mean values and
distributions. In particular, we examine the need for resummed
predictions for distributions and clarify the nomenclature and notation
used in that context, as well as the problem of matching these predictions
to fixed order computations. We then turn, in sections~\ref{sec:exp}
and \ref{sec:distributions}, to comparisons with experimental data and
the issue of non-perturbative corrections, required in order to be
able to apply parton level calculations to hadronic final state data.
We discuss the various approaches used to estimate non-perturbative
corrections, ranging from phenomenological hadronisation models to
analytical approaches based on renormalons, and shape-functions. We
also discuss methods where the standard perturbative results are
modified by use of renormalisation group improvements or the use of
`dressed gluons'. We study fits to event shape data, from
different sources, in both $e^{+}e^{-}$ annihilation and DIS and
compare the various approaches that are adopted to make theoretical
predictions.  In addition we display some of the results obtained by
fitting data for various event shape variables for parameters such as
the strong coupling, the QCD colour factors and the QCD beta function.
Lastly, in section~\ref{sec:outlook}, we present an outlook on
possible future developments in the field.

\section{Definitions and properties}
\label{sec:definitions}

We list here the most widely studied event shapes, concentrating on
those that have received significant experimental and theoretical
attention.

\subsection{$\ee$}
\emph{The} canonical event shape is the thrust~\cite{Brandt,Farhi},
$T$:
\begin{equation}
\label{eq:T}  
T = \max_{\vec n_T} \frac{ \sum_i |\vec p_i . \vec n_T|}{\sum_i
  |{\vec p}_i|}\,,
\end{equation}
where the numerator is maximised over directions of the unit vector
$\vec n_T$ and the sum is over all final-state hadron momenta $p_i$
(whose three-vectors are ${\vec p}_i$ and energies $E_i$). The
resulting $\vec n_T$ is known as the thrust axis.
In the limit of two narrow back-to-back jets $T \to 1$, while its
minimum value of $1/2$ corresponds to events with a uniform
distribution of momentum flow in all directions.  The infrared and
collinear safety of the thrust (and other event shapes) is an
essential consequence of its linearity in momenta. Often it is $1-T$
(also called $\tau$) that is referred to insofar as it is this that
vanishes in the $2$-jet limit.

A number of other commonly studied event shapes are constructed
employing the thrust axis.  Amongst these are the invariant squared
jet-mass~\cite{Clavelli} and the jet-broadening variables~\cite{CTWbroad},
defined respectively as
\begin{eqnarray}
\label{eq:rho}
\rho_{\ell,r} &=& \frac{\left(\sum_{i\in \hlr} 
  p_i \, \right)^2}{
  \left(\sum_i    E_{i}\right)^2}\,,\\
B_{\ell,r} &=& \frac{\sum_{i\in \hlr} |{\vec p}_{i}\times \vec n_T|}
{ 2\sum_i |{\vec p}_i|} \,,
\end{eqnarray}
where the plane perpendicular to the thrust axis\footnote{In the
  original definition~\cite{Clavelli}, the plane was chosen so as to
  minimise the heavy-jet mass.} is used to separate the event into
left and right hemisphere, $\hleft$ and $\hright$. Given these
definitions one can study the heavy-jet mass, $\rho_H =
\max(\rho_\ell,\rho_r)$ and wide-jet broadening, $B_W =
\max(B_\ell,B_r)$; analogously one defines the light-jet mass
($\rho_L$) and narrow-jet broadening $B_N$; finally one defines also
the sum of jet masses $\rho_S = \rho_\ell + \rho_r$ and the total jet
broadening $B_T = B_\ell + B_r$ and their differences $\rho_D = \rho_H
- \rho_L$, and $B_D = B_W - B_N$. Like $1-T$, all these variables (and
those that follow) vanish in the two-jet limit; $\rho_L$ and $B_N$ are
special in that they also vanish in the limit of three narrow jets
(three final-state partons; or in general for events with any number
of jets in the heavy hemisphere, but only one in the light hemisphere).

Another set of observables \cite{MarkJPhysRep} making use of the
thrust axis starts with the thrust major $T_M$,
\begin{equation}
T_M = 
  \max_{{\vec n}_M} \frac{ \sum_i |\vec p_i . \vec n_M|}{\sum_i
  |{\vec p}_i|}\,,\qquad\quad \vec n_M. \vec n_T=0\,,
\end{equation}
where the maximisation is performed over all directions of the unit
vector $\vec n_M$, such that $\vec n_M. \vec n_T=0$. The thrust minor,
$T_m$, is given by
\begin{equation}
  T_m =   \frac{ \sum_i |\vec p_i . \vec n_m|}{\sum_i
  |{\vec p}_i|}\,,\qquad\qquad \vec n_m = \vec n_T \times \vec n_M\,,
\end{equation}
and is sometimes also known \cite{BDMZtmin} as the (normalised)
out-of-plane momentum
$K_\mathrm{out}$. Like $\rho_L$ and $B_N$, it vanishes in the
three-jet limit (and, in general, for planar events). Finally from
$T_M$ and $T_m$ one constructs the 
oblateness, $O = T_M-T_m$.

An alternative axis is used for the spherocity $S$~\cite{Farhi},
\begin{equation}
  \label{eq:spherocity}
  S = \left(\frac{4}{\pi}\right)^2 \max_{\vec n_S}
    \left(\frac{ \sum_i |\vec p_i \times \vec n_S|}{\sum_i
      |{\vec p}_i|}\right)^2.
\end{equation}
While there exist reliable (though algorithmically slow) methods of
determining the thrust and thrust minor axes, the general properties
of the spherocity axis are less well understood and consequently the
spherocity has received less theoretical attention. An observable
similar to the thrust minor (in that it measures out-of-plane
momentum), but defined in a manner analogous to the spherocity is the
acoplanarity~\cite{Acoplanarity}, which minimises a (squared)
projection perpendicular to a plane.

It is also possible to define event shapes without reference to an
explicit axis. The best known examples are the $C$ and $D$ parameters
\cite{ERT} which are obtained from the momentum tensor
\cite{MomTensor}\footnote{We note that there is a set of earlier
  observables, the sphericity~\cite{sphericity}, planarity and
  aplanarity~\cite{planarity} based on a tensor $\Theta_{ab} = \sum_i
  p_{ai}p_{bi}$. Because of the quadratic dependence on particle
  momenta, these observable are collinear unsafe and so no longer
  widely studied.}
\begin{equation}
  \Theta_{ab} = \left(\sum_i \frac{p_{ai}p_{bi}}{|\vec p_i|} \right)
   / \sum_i
  |{\vec p}_i|\,,
\end{equation}
where $p_{ai}$ is the $a^\mathrm{th}$ component of the three vector
$\vec p_i$. In terms of the eigenvalues $\lambda_1$, $\lambda_2$ and
$\lambda_3$ of $\Theta_{ab}$, the $C$ and $D$ parameters are given by
\begin{equation}
  C = 3(\lambda_1 \lambda_2 + \lambda_2 \lambda_3 + \lambda_3
  \lambda_1)\,,\qquad\quad 
  D = 27 \lambda_1 \lambda_2 \lambda_3\,.
\end{equation}
The $C$-parameter is related also to one of a series of Fox-Wolfram
observables~\cite{FoxWolf}, $H_2$, and is sometimes equivalently
written as
\begin{equation}
  C = \frac32 \frac{\sum_{i,j} |{\vec p}_i| |{\vec p}_j| \sin^2
  \theta_{ij}}{\left(\sum_i    |{\vec p}_i|\right)^2}\,.
\end{equation}
The $D$-parameter, like the thrust minor, vanishes for all
final-states with up to 3 particles, and in general, for planar
events.

The $C$-parameter can actually be considered (in the limit of all
particles being massless) as the integral of a more
differential observable, the Energy-Energy Correlation (EEC)
\cite{Basham},
\begin{eqnarray}
  \label{eq:Sigmachi}
  \frac{d\Sigma(\chi)}{d\cos \chi} &=& \frac{d\sigma}{\sigma\, d\cos \chi}
   = \langle\, \EEC(\chi) \,\rangle\\
   \EEC(\chi) &=& \frac{\sum_{i,j} E_i E_j \delta(\cos\chi - \cos
   \theta_{ij})}
       {\left(\sum_i E_i\right)^2}
\end{eqnarray}
where the average in eq.~(\ref{eq:Sigmachi}) is carried out over all
events. 

Another set of variables that characterise the shape of the final
state are the $n$-jet resolution parameters that are generated by jet finding
algorithms. Examples of these are the JADE~\cite{JADE} and
Durham~\cite{Durham} jet clustering algorithms. One introduces
distance measures $y_{ij}$
\begin{eqnarray}
  y_{ij}^{(\mathrm{Jade})} &=&
  \frac {2 E_i E_j (1 - \cos \theta_{ij})}
        {(\sum_k E_k)^2},\\
  y_{ij}^{(\mathrm{Durham})} &=&
  \frac {2\,\mathrm{min}(E_i^2, E_j^2) (1 - \cos \theta_{ij})}
        {(\sum_k E_k)^2},
\end{eqnarray}
for each pair of particles $i$ and $j$. The pair with the smallest
$y_{ij}$ is clustered (by adding the four-momenta --- the $E$
recombination scheme) and replaced with a single pseudo-particle; the
$y_{ij}$ are recalculated and the combination procedure repeated until
all remaining $y_{ij}$ are larger than some value $y_\mathrm{cut}$.
The event-shapes based on these jet algorithms are $y_3 \equiv
y_{23}$, defined as the maximum value of $y_\mathrm{cut}$ for which
the event is clustered to 3 jets; analogously one can define $y_4$,
and so forth. Other clustering jet algorithms exist. Most differ
essentially in the definition of the distance measure $y_{ij}$ and the
recombination procedure. For example, there are E0, P and P0
variants~\cite{JADE}  
of the JADE algorithm, which differ in the details of the treatment of the
difference between energy and the modulus of the 3-momentum. The
Geneva algorithm~\cite{Geneva} is like the JADE algorithm except that
in the definition of the $y_{ij}$ it is $E_{i}+E_{j}$ that appears in the
denominator instead of the total energy. An algorithm that has been
developed and adopted recently is the Cambridge
algorithm~\cite{Cambridge}, which uses the same distance measure as
the Durham algorithm, but with a different clustering sequence.

We note that a number of variants of the above observables have
recently been introduced~\cite{Salwick}, which differ in their
treatment of massive particles.  These include the $p$-scheme where
all occurrences of $E_i$ are replaced by $|\vec p_i|$ and the
$E$-scheme where each 3-momentum $\vec p_i$ is rescaled by $E_i/|\vec
p_i|$. The former leads to a difference between the total energy in
the initial and final states, while the latter leads to a final-state
with potentially non-zero overall $3$-momentum. While such schemes do
have these small drawbacks, for certain observables, notably the jet
masses, which are quite sensitive to the masses of the hadrons (seldom
identified in experimental event-shape studies), they tend to be both
theoretically and experimentally cleaner.  Ref.~\cite{Salwick} also
introduced a `decay' scheme (an alternative is given in~\cite{GRmass})
where all hadrons are artificially decayed to massless particles.
Since a decay is by definition a stochastic process, this does not
give a unique result on an event-by-event basis, but should rather be
understood as providing a correction factor which is to be averaged
over a large ensemble of events. It is to be kept in mind that these
different schemes all lead to identical perturbative predictions (with
massless quarks) and differ only at the non-perturbative level.

A final question relating to event-shape definitions concerns the
hadron level at which measurements are made. Since shorter lived
hadrons decay during the time of flight, one has to specify whether
measurements were made at a stage before or after a given species of
hadron decays. It is important therefore when experimental results are
quoted, that they should specify which particles have been taken to be
stable and which have not.

\subsection{DIS}

As well as the $e^+e^-$ variables discussed above, it is also possible to
define, by analogy, event shapes in DIS. The frame in which DIS event
shapes can be made to most closely 
resemble those of $e^+e^-$ is the Breit frame
\cite{Breit,BreitThrust}. This is the frame in which $2x\vec P + \vec
q = 0$, where $P$ is the incoming proton momentum and $q$ the virtual
photon momentum. One defines two hemispheres, separated by the plane
normal to the photon direction: the remnant hemisphere ($\hr$,
containing the proton remnant), and the current hemisphere ($\hc$). At
the level of the quark-parton model, $\hc$ is like one hemisphere of
$\ee$ and it is therefore natural to define event shapes using only
the momenta in this
hemisphere. In contrast, any observable involving momenta in the
remnant hemisphere must take care to limit its sensitivity to the
proton remnant, whose fragmentation cannot be reliably handled within
perturbation theory. A possible alternative to studying just the current
hemisphere, is to take all particles except those in a small cone
around the proton direction~\cite{KOUTDIS}.

A feature that arises in DIS is that there are two natural choices of
axis. For example, for the thrust
\begin{equation}
  T_{nE} = \frac{\sum_{i \in \hc} |\vec p_i . \vec n|}{\sum_{i \in \hc}
  E_i}\,,
\end{equation}
one can either choose the unit vector $\vec n$ to be the photon ($z$)
axis,
$T_{zE}$, or one can choose it to be the true thrust axis, that which
maximises the sum, giving $T_{tE}$. Similarly one defines two variants
of the jet broadening,
\begin{equation}
  B_{nE} = \frac{\sum_{i \in \hc} |\vec p_{i} \times \vec n|}{2\sum_{i
  \in \hc} E_i}\,.
\end{equation}
For the jet-mass and $C$-parameter the choice of axis does not enter
into the definitions and we have
\begin{equation}
  \rho_E = \frac{(\sum_{i \in \hc} p_i)^2}{4(\sum_{i \in \hc}
  E_i)^2}\,, \qquad\quad C_E = \frac{3}{2}\frac{\sum_{i,j \in \hc}
  |\vec p_i| |\vec p_j|\sin^2\theta_{ij}}{(\sum_{i \in \hc}
  E_i)^2}\,.
\end{equation}
All the above observables can also be defined with an alternative
normalisation, $Q/2$ replacing $\sum_{i\in\hc} E_i$, in which case
they are named $T_{zQ}$ and so on. We note that $T_{zQ}$ was
originally proposed in~\cite{BreitThrust}. For a reader used to
$\ee$ event shapes the two normalisations might at first
sight seem equivalent --- however when considering a single
hemisphere, as in DIS, the equivalence is lost, and indeed there are
even events in which the current hemisphere is empty. This is a
problem for observables normalised to the sum of energies in $\hc$, to
the extent that to ensure infrared safety it is necessary to exclude
all events in which the energy present in the current hemisphere is
smaller than some not too small fraction of $Q$ (see for example
\cite{DasSalTRC}).

Additionally there are studies of
variables that vanish in the 2+1-parton limit for DIS, in particular
the $K_\mathrm{out}$ defined in analogy with the thrust minor of
$\ee$, with the thrust axis replaced by the photon axis
\cite{KOUTDIS}, or an azimuthal correlation observable~\cite{AZIMDIS}.
Rather than being examined just in $\hc$, these observables use
particles also in the remnant hemisphere (for $K_\mathrm{out}$, all
except those in a small cone around the proton).

As in $\ee$, jet rates are studied also in DIS. Unlike the
event-shapes described above, the jet-shapes make use of the momenta
in both hemispheres of the Breit frame. Their definitions are quite
similar to those of $\ee$ except that a clustering to the proton
remnant (beam jet) is also included~\cite{KTDIS}. 

\subsection{Other processes} 
Though only $\ee$ and DIS are within the scope of this review, we take
the opportunity here to note that related observables are being
considered also for other processes. Notably for Drell-Yan production
an out-of-plane momentum measurement has been proposed
in~\cite{KOUTDY} and various thrust and thrust-minor type observables
have been considered in hadron-hadron dijet production
in~\cite{Bertram:sv,Caesar,Banfi:2003nx}.

\section{Perturbative predictions}
\label{sec:perturbative}

The observables discussed above are all infrared and collinear safe
--- they do not change their value when an extra soft gluon is added
or if a parton is split into two collinear partons. As emerges from the
original discussion of 
Sterman and Weinberg~\cite{SW}, this is a necessary condition for the
cancellation of real and virtual divergences associated with such
emissions, and therefore for making finite perturbative predictions.

For an event shape that vanishes in the $n$-jet limit (which we shall
generically refer as an $n$-jet observable), the leading perturbative
contribution is of order $\as^{n-1}$. For example the thrust
distribution in $\ee$ is given by
\begin{eqnarray}
  \label{eq:thrustdist}
  \frac1\sigma \frac{d\sigma}{d(1-T)} &=& \frac{\as C_F}{2\pi}
  \left[\frac{2(3T^2 - 3T + 2)}{T(1-T)}\ln \frac{2T-1}{1-T}
    \right. \\ && \qquad \qquad \qquad\qquad
    \left. - \frac{3(3T-2)(2-T)}{1-T}
    \right] + \order{\as^2} \nonumber
\end{eqnarray}
(see e.g.~\cite{ESWBook}). 

When calculating perturbative predictions for mean values of event
shapes, as well as higher moments of their distributions, one has
integrals of the form
\begin{equation}
  \label{eq:meanV}
  \langle v^m \rangle = \int_0^{v_\mathrm{max}}
  dv \frac1\sigma \frac{d\sigma}{dv} v^m\,.
\end{equation}
We note that in general, fixed-order event-shape distributions diverge
in the limit as $v$ goes to $0$, cf.\ eq.~(\ref{eq:thrustdist}) for
$(1-T) \to 0$. In eq.~(\ref{eq:meanV}), the weighting with a power of
$v$ is sufficient to render the singularity integrable, and the
integral is dominated by large $v$ (and so large transverse momenta of
order $Q$, the hard scale). This dominance of a single scale ensures
that the coefficients in the perturbative expansion for $\langle v^m
\rangle$ are well-behaved (for example they are free of any
enhancements associated with logarithms of ratios of disparate
scales).

Beyond leading order, perturbative calculations involve complex
cancellations between soft and collinear real and virtual
contributions and are nearly always left to general-purpose
``fixed-order Monte Carlo'' programs. The current state of the art is
next-to-leading order (NLO), and available programs include: for $\ee$
to 3 jets, EVENT~\cite{EVENT}, EERAD~\cite{EERAD} and EVENT2
\cite{EVENT2}; for $\ee$ to 4 jets, Menlo Parc~\cite{MENLO}, Mercutio
\cite{Mercutio} and EERAD2~\cite{EERAD2}; for DIS to $2+1$ jets,
MEPJET~\cite{MEPJET}, DISENT~\cite{EVENT2} and DISASTER++
\cite{Disaster}; and for all the above processes and additionally DIS
to $3+1$ jets and various hadron-hadron and photo-production processes,
NLOJET++~\cite{NLOJET,NLOJETHH} (for photo-production, there is
additionally JETVIP~\cite{JETVIP}). In recent years much progress has
been made towards NNLO calculations, though complete results
remain to be obtained (for a review of the current situation, see
\cite{GehrmannReview}).

For the event shape distributions themselves, however, fixed order
perturbative estimates are only of use away from the $v=0$ region, due
to the singular behaviour of the fixed order coefficients in the $v
\to 0$ limit. In this limit, at higher orders each power of $\as$ is
accompanied by a coefficient which grows as $\ln^2 v$. These problems
arise because when $v$ is small one places a restriction on real
emissions without any corresponding restriction on virtual
contributions --- the resulting incompleteness of cancellations
between logarithmically divergent real and virtual contributions is
the origin of the order-by-order logarithmic enhancement of the
perturbative contributions. To obtain a meaningful answer it is
therefore necessary to perform an all-orders \emph{resummation} of
logarithmically enhanced terms.

\subsection{Resummation}

When discussing resummations it is convenient to refer to the
integrated distribution,
\begin{equation}
  \label{eq:intdist}
  R(v) = \int_0^v dv' \frac1\sigma \frac{d\sigma}{dv'}\,.
\end{equation}
Quite generally $R(v)$ has a perturbative expansion of the form
\begin{equation}
  \label{eq:Dumbsum}
  R(v) = 1 + \sum_{n=1} \left(\frac{\as}{2\pi}\right)^n
    \left(\sum_{m=0}^{2n} R_{nm} \ln^m 
    \frac1v + \order{v}\right)\,.
\end{equation}
One convention is to refer to all terms $\as^n \ln^{2n} 1/v$ as
leading logarithms (LL), terms $\as^n \ln^{2n-1} 1/v$ as
next-to-leading logarithms (NLL), etc., and within this hierarchy a
resummation may account for all LL terms, or all LL and NLL terms and
so forth. Such a resummation gives a convergent answer up to values of
$L \equiv \ln 1/v \sim \as^{-1/2}$, beyond which terms that are
formally subleading can become as important as the leading terms (for
example if $L \sim 1/\as$, then the NNLL term $\as^3L^4$ is of the
same order as the LL term $\as L^2$). At its limit of validity, $L
\sim \as^{-1/2}$, a N$^p$LL resummation neglects terms of relative
accuracy $\as^{(p+1)/2}$.

An important point though is that for nearly all observables
that have been resummed there exists a property of exponentiation:
\begin{equation}
  \label{eq:ExpResum}
  R(v) = \exp\left( \sum_{n=1} \left(\frac{\as}{2\pi}\right)^n
    \left(\sum_{m=0}^{n+1} G_{nm} \ln^m
    \frac1v + \order{v}\right) \right).
\end{equation}
In some cases $R$ is written in terms of a sum of such exponentiated
contributions; in certain other cases the exponentiation holds only
for a suitable integral (e.g.\ Fourier) transform of the observable.
The fundamental point of exponentiation is that the inner sum, over
$m$, now runs only up to $n+1$, instead of $2n$ as was the case in
eq.~(\ref{eq:Dumbsum}). With the exponentiated form for the
resummation, the nomenclature ``(next-to)-leading-logarithmic''
acquires a different meaning --- N$^p$LL now refers to all terms in
the exponent $\as^n L^{n+1-p}$. To distinguish the two classification
schemes we refer to them as N$^p$LL$_{R}$ and N$^p$LL$_{\ln R}$.

The crucial difference between N$^p$LL$_{R}$ and N$^p$LL$_{\ln R}$
resummations lies in the range of validity and their accuracy. A
N$^p$LL$_{\ln R}$ resummation remains convergent considerably further
in $L$, up to $L \sim 1/\as$ (corresponding usually to the peak of the
distribution $dR/dv$); at this limit, neglected terms are of relative
order $\as^p$. Consequently a N$^p$LL$_{\ln R}$ resummation includes
not only all N$^p$LL$_{R}$ terms, but also considerably more. From now
when we use the term N$^p$LL it should be taken to mean N$^p$LL$_{\ln
  R}$.

Two main ingredients are involved in the resummations as shown in
eq.~(\ref{eq:ExpResum}). Firstly one needs to find a way of writing
the observable as a factorised expression of terms for individual
emissions. This is often achieved with the aid of one or more 
integral (Mellin,
Fourier) transforms. Secondly one approximates the
multi-emission matrix element as a product of individual independent
emission factors.\footnote{There exist cases in which, at NLL
accuracy, this is not quite sufficient, specifically for observables
that are referred to as `non-global'~\cite{NGOneJet}. This will be
discussed below.}

It is to be kept in mind that not all observables exponentiate. The
JADE jet resolution is the best known such example~\cite{JadeNoExp}
(and it has not yet been resummed).  The general criteria for
exponentiation are discussed in~\cite{Caesar}.

Resummed results to NLL accuracy exist for a number of $2$-jet observables in
$\ee \to 2$ jets: the thrust~\cite{CTTWthrust,CTTWlong}, the heavy-jet
mass~\cite{CTTWmass,CTTWlong} and the single-jet and light-jet masses
\cite{BurbyGlover,NGOneJet}, jet broadenings
\cite{CTWbroad,DLMSBroad,NGOneJet}, $C$-parameter~\cite{CWcparam},
Durham and Cambridge jet resolutions
\cite{Durham,JetsDisSchmell,NumSum}, the thrust major and oblateness
\cite{NumSum}, EEC~\cite{EECearly,EECresum} and
event-shape/energy-flow correlations~\cite{BKScorrel}; for $3$-jet
observables in $\ee \to 3$
jet events we have the thrust minor~\cite{BDMZtmin} and the $D$-parameter
\cite{BDMZdpar}. For DIS to $1+1$ jets the current-hemisphere
observables are resummed in refs.~\cite{ADS,DasSalBroad,DasSalTRC} and
the jet rates in~\cite{KTDIS} (though only to NLL$_{R}$ accuracy)
while the $2+1$ jet observables, $K_\mathrm{out}$ and azimuthal
correlations, have been resummed in~\cite{KOUTDIS,AZIMDIS}. Methods
for automated resummation of arbitrary observables are currently in
development for a range of processes\cite{Caesar}, and techniques have
also being developed for dealing with arbitrary processes~\cite{BCMN}.
We also note that for the $\ee$ thrust and
heavy-jet mass there have been investigations of certain classes
of corrections beyond NLL accuracy~\cite{GRthrust,GRmass,GardMan}. The above
resummations all apply to events with only light
quarks. Investigations to NLL$_R$ accuracy have also been performed
for $\ee$ jet rates with heavy quarks~\cite{KraussRodrigo}.

The above resummations are all for $n$-jet observables in
the $n$-jet limit. For some $3$-jet observables it is useful to have
resummations in the $2$-jet limit (because it is here that most of the
data lie), though currently little attention has been devoted to such
resummations, with just NLL$_R$ results for jet-resolution parameters
\cite{Durham,KTDIS} and NLL$_{\ln R}$ results for the light-jet mass
and narrow-jet broadening~\cite{BurbyGlover,NGOneJet}. Furthermore in
addition to the resummation of the $C$-parameter for $C\to0$ there
exists a LL resummation~\cite{Catani:1997xc} (that could
straightforward be extended to NLL) of a `shoulder' structure at
$C=3/4$ (related to a step function at LO), which corresponds to
symmetric three jet events.

Some of the observables (for example many single-jet observables) have
the property that they are sensitive to emissions only in some of the
phase space. These are referred to as non-global observables and NLL
resummed predictions for them require that one account for coherent
ensembles of energy-ordered large-angle gluons. This has so far been
done only in the large-$N_C$ limit~\cite{NGOneJet,EnergyFlows,
  ApplebySeymour,DasSalTRC, BKScorrel,DMNonGlobal} (for reviews see
\cite{NGreviews}), though progress is being made in extending this to
finite $N_C$~\cite{Weigert:2003mm}.

\begin{figure}
  \begin{center}
    \includegraphics[width=0.6\textwidth]{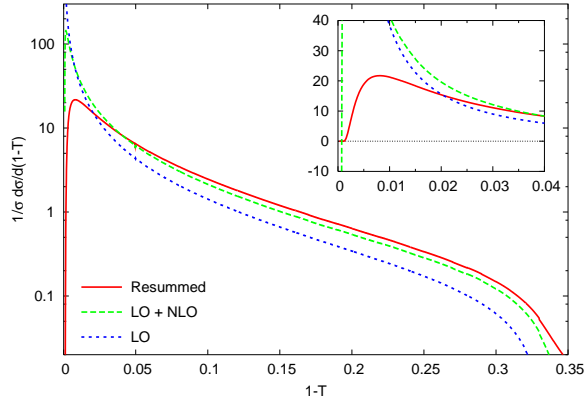}
  \end{center}
  \caption{Comparison of the fixed-order (LO, NLO) and resummed (NLL)
    predictions for the thrust distribution in $\ee$ at $Q=M_Z$. The
    resummed distribution has been matched ($\ln R$ scheme, see below)
    so as to include the full LO and NLO contributions. The inset
    differs from the main plot only in the scales.}
  \label{fig:thrLONLO}
\end{figure}

To illustrate the impact of resummations we show in
figure~\ref{fig:thrLONLO} the LO, NLO and resummed results for the
thrust distribution. Not only is the divergence in the LO
distribution clearly visible, but one also observes a marked
difference in the qualitative behaviour of the LO and NLO results,
both of which show large differences relative to the NLL resummed
prediction at small $1-T$.

For the practical use of resummed results an important step is that of
matching with the fixed-order prediction. At the simplest level one
may think of matching as simply adding the fixed-order and resummed
predictions while subtracting out the doubly counted (logarithmic)
terms. Such a procedure is however too naive in that the fixed-order
contribution generally contains terms that are subleading with respect
to the terms included in the resummation, but which still diverge. For
example when matching NLL and NLO calculations, one finds that the NLO
result has a term $\sim R_{21} \as^2 L$, which in the distribution
diverges as $\as^2/v$. This is unphysical and the matching procedure must
be sufficiently sophisticated so as to avoid this problem, and in fact it 
should nearly always ensure that the matched distribution goes to zero
for $v\to0$.  Several procedures exist, such as $\ln R$ and $R$
matching~\cite{CTTWlong} and multiplicative matching
\cite{DasSalBroad} and they differ from one-another only in terms that
are NNLO and NNLL, i.e.\ formally beyond the state-of-the-art
accuracy. We note the matching generally ensures that the resummation,
in addition to being correct at NLL accuracy (our shorthand for
NLL$_{\ln R}$), is also correct at NNLL$_R$ accuracy. Matching also
involves certain other subtleties, for example the `modification of
the logarithm', whereby $\ln 1/v$ is replaced with $\ln (1 + 1/v -
1/v_{\max})$, where $v_{\max}$ is the largest kinematically allowed
value for the observable. This ensures that the logarithms go to zero
at $v_{\max}$, rather than at some arbitrary
point (usually $v=1$), which is necessary in order for the resulting
matched distribution to vanish at $v_{\max}$.

\section{Mean values, hadronisation corrections and comparisons to experiment }
\label{sec:exp}
\label{sec:means}
\label{sec:meanPC}

Much of the interest in event shapes stems from the wealth of data
that is available, covering a large range of centre of mass energies
($\ee$) or photon virtualities (DIS). The $\ee$ data comes from the
pre-LEP experiments~\cite{OldEEResults}, the four LEP experiments
\cite{
Decamp:1990cg,
Decamp:1992wz,
Buskulic:1996tt,Barate:1996fi,
AlephConf2000027,Aleph03}
\cite{
Abreu:1996mk,
Abreu:1999rc,
Abreu:2000ck,
Abdallah:2002xz}
\cite{
Adriani:1992gs,
Adeva:1992gv,
Adriani:1993gk,
Acciarri:1997dn,
Acciarri:1995ia,
Acciarri:1997xr,
Acciarri:1998gz,
Acciarri:2000hm,
Achard:2002kv}
\cite{
  Acton:1992fa,
  Acton:1993zh, 
  Alexander:1996kh, 
  Ackerstaff:1997kk, 
  Abbiendi:1999sx} %
and SLD~\cite{Abe:1994mf}. In addition, because many observables have
been proposed only in the last ten or so years and owing to the
particular interest (see below) in data at moderate centre of mass
energies, the
JADE data have been re-analysed %
\cite{MovillaFernandez:1997fr,
  Biebel:1999zt,
  Pfeifenschneider:1999rz,
  MovillaFernandez:2001ed,%
  Kluth:2003uq}.%
\footnote{Data below $M_Z$ have also been obtained from LEP~1
  \cite{Acciarri:1997dn,Abdallah:2002xz,OpalRadiative} by considering
  events with an isolated hard final-state photon and treating them as
  if they were pure QCD events whose centre of mass energy is that of
  the recoiling hadronic system. Such a procedure is untrustworthy
  because it assumes that one can factorise the gluon production from
  the photon production. This is only the case when there is strong
  ordering in transverse momenta between the photon and gluon, and is
  not therefore applicable when both the photon and the gluon are
  hard. Tests with Monte Carlo event generators which may indicate
  that any `non-factorisation' is small~\cite{Acciarri:1997dn} are not
  reliable, because the event generator used almost certainly does not
  contain the full 1-photon, 1-gluon matrix element. It is also to be
  noted that the isolation cuts on the photon will bias the
  distribution of the event shapes, in some cases
  \cite{Acciarri:1997dn,Abdallah:2002xz} similarly to an event-shape
  energy-flow correlation~\cite{BKScorrel,DMNonGlobal}.  Therefore we
  would argue that before relying on these data, one should at the
  very least compare the factorisation approximation with exact LO
  calculations, which can be straightforwardly obtained using packages
  such as Grace~\cite{Grace}, CompHEP~\cite{CompHEP} or Amegic++
  \cite{Amegic}.} %
Results in DIS come from both H1
\cite{Adloff:1997gq,Adloff:2000,Martyn:2000jk} (mean values and
distributions) and ZEUS~\cite{ZEUSmeans} (means), \cite{ZEUSRapGap}
(means and distributions in rapidity gap events).

\begin{figure}
  \begin{center}
    \includegraphics[width=0.48\textwidth]{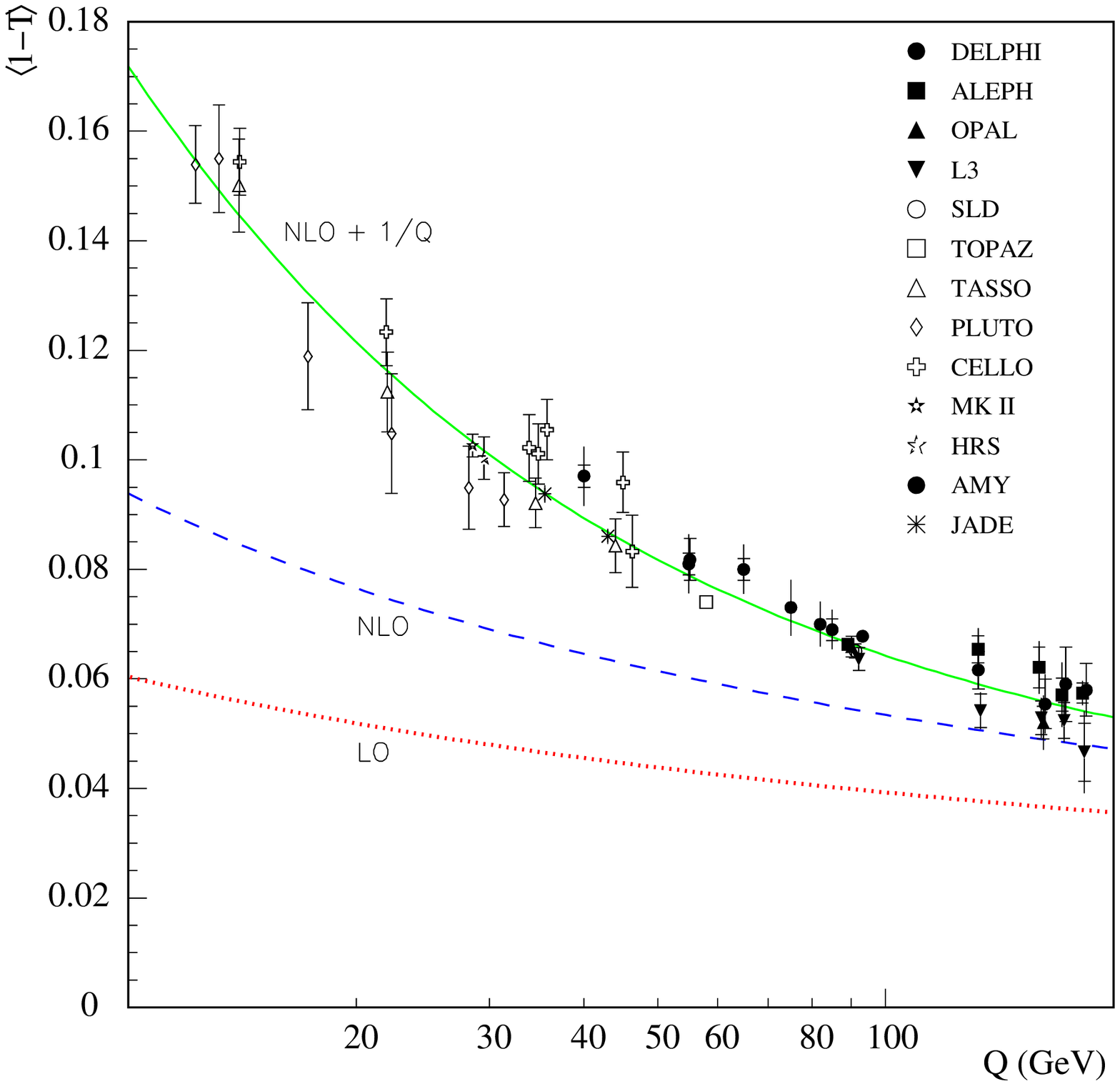}\hfill
    \includegraphics[width=0.48\textwidth]{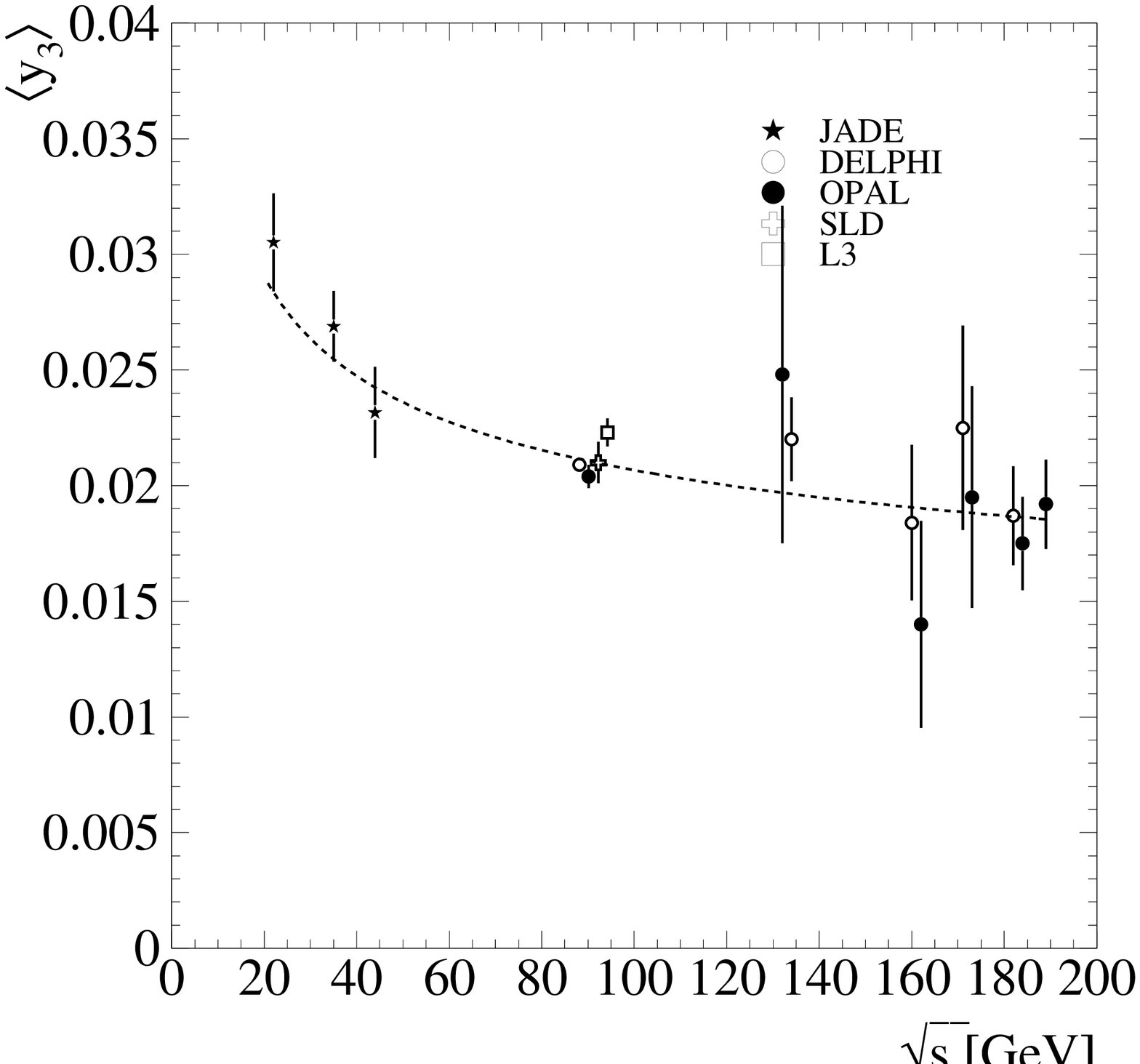}
  \end{center}
  \caption{Left: data for the mean value of the $\ee$ thrust, compared to
    LO, NLO and NLO+$\order{\Lambda/Q}$;
    right: data for the Durham jet resolution $y_{23}$ compared to the
    NLO prediction (figure taken from
    \cite{MovillaFernandez:2001ed}).}
  \label{fig:meandata}
\end{figure}


Let us start our discussion of the data by examining mean values
(distributions are left until section~\ref{sec:distributions}).  As we
have already mentioned, one of the most appealing properties of event
shapes, in terms of testing QCD, is the fact that they are calculable
in perturbation theory and so provide a direct method for the
extraction of $\alpha_s$ as well as testing other parameters of the
theory such as the colour factors $C_F$ and $C_A$ by using a wealth of
available experimental data. However one obstruction to the clean
extraction of these fundamental parameters is the presence in many
cases, of significant non-perturbative effects that typically fall as
inverse powers $\propto 1/Q^{2p}$ of $Q$, the hard scale in the reaction
(the centre-of--mass energy in $e^{+}e^{-}$ annihilation and the
momentum transfer in DIS).  The importance of such {\it{power
    corrections}} varies from observable to observable. For example,
it can be seen from Fig.~\ref{fig:meandata} (left) that the data for
the mean value of the thrust variable need a significant component
$\sim \Lambda/Q$ in addition to the LO and NLO fixed order
perturbative estimates, in order to be described. On the other
hand the comparison (right) for the mean value of the Durham jet
resolution parameter $y_{23}$ with the NLO prediction alone is
satisfactory, without the need for any substantial power correction
term. The problem of non-perturbative corrections is so fundamental in
event-shape studies that it is worthwhile giving a brief overview of
the main approaches that are used.

\subsection{Theoretical approaches to hadronisation corrections}
\label{sec:meanhadronisation}

No statement can be made with standard perturbative methods about
the size of these power corrections and hence the earliest methods
adopted to quantify them were phenomenological hadronisation models
embedded in Monte Carlo event generators~\cite{Herwig,Pythia,Ariadne}.
One of the main issues involved in using such hadronisation models is
the existence of several adjustable parameters. This essentially means
that a satisfactory description of data can be obtained by tuning the
parameters, which does not allow much insight into the physical origin
of such power behaved terms which are themselves of intrinsic
theoretical interest (for a more detailed critique of hadronisation
models see~\cite{Webfrasc}).

However since the mid-nineties there have been developments that have
had a significant impact on the theoretical understanding of power
corrections.  Perhaps the most popular method of estimating power
corrections is based on the renormalon model.  In this approach one
examines high-order terms $\as^n$ of the perturbative series that are
enhanced by coefficients $b^n n!$, where $b$ is the first coefficient
of the $\beta$ function. The factorial divergence leads to an
ambiguity in the sum of the series of order $(\Lambda/Q)^2p$, where
the value of $p$ depends on the speed with which the high-order terms
diverge. This approach has seen far more applications than can
possibly be described here and for a full discussion and further
references on renormalons the reader is referred to
\cite{Beneke:1998ui}.  The most extensive phenomenological
applications of renormalon based ideas have been in the estimation of
power corrections for event-shape variables. The reason is that event
shapes have much larger (and so experimentally more visible) power
corrections (typically $\Lambda/Q$) than most other observables
(typically $(\Lambda/Q)^2$ or smaller).

For event shape variables in $e^{+}e^{-}$ annihilation in the two-jet
limit, renormalon-inspired (or related) studies of power corrections
have been carried out in refs.~\cite{ ManWise1, korch1, korch2,
  korch3, korch4, korch5, BenBrau, web1, dok1, dok2, dok3, EECresum,
  Akh1, Akh2, mil1, mil2, mil3, mil4, mil5, Salwick, GardGrun,
  GRthrust, GRmass, Gard, GardMan, BergSterm, ManWise2,
  ManWise3,Trocsanyi:1999ta}. For the $1+1$ jet limit of DIS, results
on power corrections based on the renormalon approach can be found in
\cite{DasWeb1,DasWebmil}. Studies in the $3$-jet limit have been given
in~\cite{KOUTDIS,AZIMDIS,KOUTDY,BDMZtmin,BDMZdpar,LAPPSE}.

\subsubsection{Dokshitzer-Webber approach.}
\label{sec:DWdetails}

The approach that has been most commonly used in comparing theoretical
predictions with experimental data is that initiated by Dokshitzer and
Webber~\cite{web1,dok1,Webber:1995ka,dok2,dok3,EECresum,
  mil1,mil2,mil3,mil4, mil5,DasWeb1,DasWebmil}.  Here the full result
for the mean value, $\langle v\rangle$, is given by
\begin{equation}
  \label{eq:meancorrections}
  \langle v\rangle = \langle v\rangle_\mathrm{PT} + C_V \cP_p\,,
\end{equation}
where $\langle v\rangle_\mathrm{PT}$ is the perturbative prediction
for the mean value. The hadronisation correction is included through
the term $C_V \cP_p$, where $p$ is related to the speed of divergence of
the renormalon series (as discussed above), $C_V$ is an
observable-dependent (calculable) coefficient and $\cP_p$ is a
non-perturbative factor, scaling as $(\Lambda/Q)^{2p}$ which is
hypothesised to be common across a whole class of observables with the
same value of $p$ (strictly speaking the full story is a little more
complicated, see e.g.\ \cite{dok2}). Most event shapes have $p=1/2$,
implying a leading power correction scaling as $\Lambda/Q$,
essentially a consequence of the observables' linearity on soft
momenta. For these observables with $p=1/2$ the form for $\cP \equiv
\cP_{1/2}$ that has become standard is~\cite{dok1,mil2},
\begin{equation}
  \label{eq:cPfin}
  \cP\> \equiv\>  \frac{4C_F}{\pi^2}\cM \frac{\mu_I}{Q}
  \left\{ \alpha_0(\mu_I)- \as(\mu_R)
    -\beta_0\frac{\as^2}{2\pi}\left(\ln\frac{\mu_R}{\mu_I} 
      +\frac{K}{\beta_0}+1\right) \right\}\>,
\end{equation}
where $\as\equiv \alpha_{\MSbar}(\mu_R)$, $\beta_0 =\frac{11}{3} C_A
-\frac{2}{3}n_f$ and $K = C_A(67/18-\pi^2/6) - 5n_f/9$. In
eq.~(\ref{eq:cPfin}) an 
arbitrary \emph{infrared matching scale} $\mu_I$ has been introduced,
intended to separate the perturbative and non-perturbative regions. It
is usually taken to be $2$~GeV (and for systematic error estimates it
is varied between $1$ and $3$~GeV). The only truly non-perturbative
ingredient in eq.~(\ref{eq:cPfin}) is $\alpha_0(\mu_I)$, which can be
interpreted as the average value of an infrared finite strong coupling
for scales below $\mu_I$ (such a concept was first applied
phenomenologically in~\cite{YuriErice}). Though one could imagine
estimating it from 
lattice studies of the coupling (such as~\cite{Boucaud:2001un}), one
should keep in mind that the coupling in the infrared is not a
uniquely defined quantity. In practice the phenomenological test of
the renormalon approach will be that $\alpha_0$ has a consistent value
across all observables (in the $p=1/2$ class).

The terms with negative sign in the parenthesis in
eq.~(\ref{eq:cPfin}) are a consequence of merging standard NLO
perturbative results, which include the small spurious contributions
from the infrared region (scales up to $\mu_I$), with the
non-perturbative power correction that accounts correctly for scales
from zero to $\mu_I$.  Carrying out the subtraction of the
perturbative terms that arise from the infrared region, below $\mu_I$,
amounts to inclusion of the negative sign terms in parenthesis above.
Note that this subtraction procedure must be carried out to the level
of accuracy of the corresponding perturbative estimate. In other words
if a perturbative estimate becomes available at $\mathcal{O}\left
  ({\alpha_s^3} \right)$ then an additional subtraction term
proportional to $\alpha_s^3$ is required.  With the subtraction
procedure carried out to $\mathcal{O}(\alpha_s^2)$, as above, one
expects a residual $\mu_I$ dependence $\sim
\mathcal{O}\left({\alpha_s}^3 (\mu_I)\right)$.  The scale $\mu_R$ is
the renormalisation scale, which should be taken of the same order as
the hard scale $Q$.

The `Milan factor', $\cM$ in eq.~(\ref{eq:cPfin}), is
\cite{mil1,mil2,mil4,mil5,DasWebmil}
\begin{equation}
  \label{eq:milanfactor}
  \cM = \frac3{64} \frac{(128\pi (1+\ln 2) - 35 \pi^2)C_A -
    10\pi^2 T_R n_f)}{11C_A - 4 T_R n_f} \simeq 1.49\,,
\end{equation}
where the numerical result is given for $n_f=3$, since only light
flavours will be active at the relevant (low) scales. It
%
%
accounts for the fact that the usual, `naive', calculation for $C_V$
is carried out on the basis of a decaying virtual gluon (i.e.\ cutting
a bubble in the chain of vacuum polarisation insertions that lead to
the running of the coupling), but without fully accounting for
the non-inclusiveness of that decay when dealing with the observable.
It was pointed out in~\cite{Nassey} that this is inconsistent and full
(two-loop) calculations revealed~\cite{mil1,mil2,DasWebmil} that if
the `naive' $C_V$ coefficient is calculated in an appropriate scheme,
then the factor $\cM$ comes out to be universal. As discussed in
\cite{mil2} this is essentially a consequence of the fact that
regardless of whether one accounts for the virtual gluon decay, the
distribution of `non-perturbative' transverse momentum is independent
of rapidity (the `tube-model' of~\cite{FeynmanTube,Webber94Tube},
based essentially on boost invariance).

While the factor $\cM$ corrects the $C_V$ at two-loop level, the
question of even higher order corrections is still open, although one
can argue that such effects will be suppressed by a factor $\sim
\alpha_0(\mu_I)/\pi$ relative to the leading power correction.  This
argument relies on the hope that the strong coupling remains moderate
even in the infrared region.

\begin{table}[b]
\begin{center}
  \begin{tabular}{|c||c|c|c|c|c|c|}
\hline
$V$ & $\tau$ & $\rho$ & $\rho_h$ & $C$  & $B_T$ & $B_W$   \\ \hline
$C_V$ & $2$  &  $1$   &   $1$    & $3\pi$ & 
 $\frac{\pi}{2\sqrt{C_F \alpha_s}} - \frac{\beta_0}{6C_F} + \bar\eta_0$ 
& $\frac{\pi}{4\sqrt{2C_F \alpha_s}}- \frac{\beta_0}{24C_F} + \frac{\bar\eta_0}{2}$
\\ \hline
  \end{tabular}
\end{center}
  \caption{coefficients of $1/Q$ power corrections for $e^{+}e^{-}$
    event shapes; $\bar\eta_0 \simeq 0.136$.}
\label{tab:cvs}
\end{table}

\begin{table}[b]
\begin{center}
  \begin{tabular}{|c||c|c|c|c|c|c|}
\hline
$V$ & $\tau_{tE}$ & $\tau_{zE}$ & $\rho_E$ & $C_E$  & $B_{tE}$ & $B_{zE}$    \\ \hline
$C_V$ & $2$  &  $2$   &   $1$    & $3\pi$ & 
 $\frac{\pi}{2\sqrt{C_F \alpha_s}} - \frac{\beta_0}{6C_F} + \bar\eta_0$ 
& 
\mbox{\begin{minipage}[c]{0.25\textwidth}
    \centering
    $\frac{\pi}{2\sqrt{2C_F \alpha_s}}- \frac{\beta_0}{12 C_F}+$\\$
    +\frac{\pi}{2C_F\alpha_s} \frac{d\ln q}{d \ln Q^2}+ \bar\eta_0$
\end{minipage}}
\\ \hline
  \end{tabular}
\end{center}
  \caption{coefficients of $1/Q$ power corrections for DIS event
    shapes; notation as in table~\ref{tab:cvs} and additionally
    $q(x,Q^2) = \sum_j e_j^2 (q_j(x,Q^2)+\bar{q}_j(x,Q^2))$, with the
    sum being over quark flavours. } 
\label{tab:cvsDIS}
\end{table}

Tables 1 and 2 display the values of the $C_V$ coefficients obtained
in the 2-jet $e^{+}e^{-}$ and and $1+1$-jet DIS cases respectively.
Note the different behaviour, proportional to $1/\sqrt{\alpha_s}$,
that arises in the case of the jet broadenings in both processes, a
consequence of the fact that the broadening is more sensitive to
recoil induced by perturbative radiation~\cite{mil3}. The improved
theoretical understanding that led to this prediction was strongly
stimulated by experimental analysis~\cite{MovillaFernandez:1998mr} of
features of earlier predictions~\cite{Webber:1995ka,mil2} that were
inconsistent with the data. Since the origin of the
$1/\sqrt{\alpha_s}$ enhancement is a perturbative, non-perturbative
interplay best explained with reference to the power correction for
the distribution, we delay its discussion to
section~\ref{sec:anldist}.

Another consequence of this interplay of the broadening power
correction with perturbative radiation is the fact that the DIS
broadening $B_{zE}$ has a Bjorken $x$ dependent term corresponding to
DGLAP evolution of the parton densities, as indicated by the term
proportional to $d\ln q(x)/d \ln Q^2$~\cite{DasSalBroad}, where $q(x)$
denotes a parton density function.  The coefficients for the other DIS
variables are $x$ independent since in those cases the power
correction arises from soft emission alone.

We also mention the result for the power correction to another
interesting variable, the energy-energy correlation (EEC), in the
back-to-back region.  Like the broadening this variable exhibits the
impact of perturbative-non--perturbative interplay. The results are
fractional power corrections that vary as $1/Q^{0.37}$ (for the
quark-gluon correlation) and $1/Q^{0.74}$ for the hard quark-antiquark
correlation~\cite{EECresum}, instead of $1/Q$ and $\ln Q^2/Q^2$
contributions respectively, that would be obtained by considering NP
emission without the presence of harder perturbative emissions.

Observables for which the power corrections have yet to be understood
include the jet resolution parameters. Renormalon based predictions
were given in~\cite{Webber:1995ka} suggesting a $\Lambda/Q$ correction
for $\langle y_{23}^\mathrm{JADE}\rangle$ and $(\Lambda/Q)^2$ for
$\langle y_{23}^\mathrm{Durham}\rangle$. However it seems that in both
cases there could be significant perturbative non-perturbative
interplay which will complicate the picture. Nevertheless, as we have
seen in fig.~\ref{fig:meandata} (right) any correction to $\langle
y_{23}^\mathrm{Durham}\rangle$ is certainly small compared to that for
other observables.

A general point to be kept in mind is that the universality of
$\alpha_0$ can be broken by contributions of order $\Lambda^2/m_h Q
\sim \Lambda/Q$, where $m_h$ is the mass scale for hadrons
\cite{Salwick}. Whether this happens or not depends on the hadron mass
scheme in which the observable is defined (cf.\ 
section~\ref{sec:definitions}). Most observables are implicitly in the
$p$-scheme, which generally involves a small negative breaking of
universality (which is almost observable-independent, so that an
illusion of universality will persist). The jet masses are an
exception and in their usual definitions they have a significant
(positive) universality breaking component. The $E$-scheme is free of
universality breaking terms, as is the decay scheme (since hadron
masses are zero). We also note that the power correction associated
with hadron mass effects is enhanced by an anomalous dimension, $(\ln
Q)^A$ with $A=4C_A/\beta_0\simeq 1.6$. The normal `renormalon' power
correction would also be expected to have an anomalous dimension, but
it has yet to be calculated.

\subsubsection{Higher moments of two-jet observables.}
\label{sec:moments}

As well as studies of mean values (first moments) of observables there
has also been work on higher-moments, $\langle v^m \rangle$ with
$m>1$. Simple renormalon-inspired arguments suggest that the
$m^\mathrm{th}$ moments of event shapes should have as their leading
power correction at most a $(\Lambda/Q)^m$ contribution (essentially
since $v^m$ vanishes at $(k_t/Q)^m$ for small $k_t$). However as was
pointed out by Webber~\cite{Webber:1997zj}, the fact (to be discussed
in section~\ref{sec:anldist}) that the $\Lambda/Q$ power correction
essentially corresponds to a shift of the distribution (for $T$, $C$
and $\rho_H$, but not for the broadenings), means that to a first
approximation we will have
\begin{equation}
  \label{eq:momentPC}
  \langle v^m \rangle = \langle v^m \rangle_\mathrm{PT} + m\, \langle
  v^{m-1} \rangle_\mathrm{PT}\, C_V \cP + \ldots\,.
\end{equation}
Given that $\langle v^{m-1} \rangle_\mathrm{PT}$ is of order $\as$,
all higher moments of event shapes will receive power corrections of
order $\as \Lambda/Q$, which is parametrically larger than
$(\Lambda/Q)^m$. Strictly speaking it is not possible to guarantee the
coefficient of $\as \Lambda/Q$ given in eq.~(\ref{eq:momentPC}), since
the shift approximation on which it is based holds only in the $2$-jet
region, whereas the dominant contribution to the relevant integral
comes from the $3$-jet region. For the $C$-parameter an alternative
coefficient for the $\as \Lambda/Q$ power correction has been proposed
in~\cite{korch4}, however since it also is not  based on a full
calculation of the power correction in the three-jet region (which
does not yet exist), it is subject to precisely the same reservations
as eq.~(\ref{eq:momentPC}).

We note also an interesting result by Gardi~\cite{Gard} regarding the
exact renormalon analysis of $\langle (1-T)^2 \rangle$. While it is
clear that the physical answer $\as \Lambda/Q$ will not appear in a
leading renormalon analysis (the extra factor of $\as$ means that it
is associated with a subleading renormalon) the calculation
\cite{Gard} has the surprising result that the leading renormalon
contribution is not $(\Lambda/Q)^2$ as naively expected but rather
$(\Lambda/Q)^3$.

\subsubsection{Power corrections to three-jet observables.}
\label{eq:threejet}

Considerable progress has been made in recent years in the calculation
of power corrections for three-jet observables (those that vanish in
the two and the three-jet limits). In the three-jet limit, there exist
explicit results for the thrust minor~\cite{BDMZtmin} and the
$D$-parameter~\cite{BDMZdpar} in $\ee\to 3\;\mathrm{jets}$, the out of
plane momentum in Drell-Yan plus jet~\cite{KOUTDY} and in $2+1$-jet
DIS~\cite{KOUTDIS}, as well as azimuthal correlations in DIS
\cite{AZIMDIS}. Except for the $D$-parameter, all these calculations
involve perturbative, non-perturbative interplay in a manner similar
to that of the broadenings and back-to-back EEC. The explicit forms
for the results are rather complicated and so we refer the reader to
the original publications.

Simpler (though numerical) results have been obtained~\cite{LAPPSE} in
the case of the $D$-parameter (confirmed also by
\cite{BDMZdpar,GiuliaPrivate}) integrated over all 3-jet
configurations, where one finds
\begin{equation}
  \label{eq:Dpower}
  C_D \,\simeq\, \left(118.0 + 34.7
    \frac{C_A}{C_F}\right)\frac{\as(\mu_R)}{2\pi} \,\simeq\, 196.0
  \frac{\as(\mu_R)}{2\pi} \,. 
\end{equation}
The power of $\as$ comes from the matrix-element weighting of the
$3$-jet configurations and the fact that the $D$-parameter vanishes in
the $2$-jet limit.

A final $3$-jet result of interest~\cite{LAPPSE} is that for the
$C$-parameter just above the Sudakov shoulder, at $C=3/4$. This is the
only case where a proper calculation exists for the power correction
to a two-jet observable in the three jet limit:
\begin{equation}
  \label{eq:C34power}
  C_{C@\frac34} \,\simeq\, 2.110\left(1 + \frac{C_A}{2C_F}\right)
  \,\simeq\, 4.484\,. 
\end{equation}
This is slightly less than half the correction that appears in the
two-jet limit ($C_C=3\pi$).

A point that emerges clearly from eqs.~(\ref{eq:Dpower}) and
(\ref{eq:C34power}), but that is relevant for all $3$-jet studies is
that the power correction acquires a dependence on $C_A$, i.e.\ there
is sensitivity to hadronisation from a gluon. In situations where one
selects only three-jet events there is additionally non trivial
dependence on the geometry, due to the coherence between the three
jets. Comparisons to data for such observables would therefore allow a
powerful test of the renormalon-inspired picture. In particular, other
models, such as the flux tube model~\cite{FeynmanTube,Webber94Tube}
(based essentially on boost invariance along the $q\bar q$ axis),
which in the $2$-jet limit give the same predictions as
renormalon-inspired approaches, cannot as naturally be extended to the
$3$-jet case.

\subsection{Fits to data}

One of the most widespread ways of testing the Dokshitzer-Webber
approach to hadronisation is to carry out simultaneous fits to the
mean value data for $\alpha_s$ and $\alpha_0$.
Figure~\ref{fig:meandata}, where the term labelled $1/Q$ is actually
of the form $C_V \cP$, shows the good agreement that is obtained with
the thrust data. The quality of agreement is similar for other
observables (see for example fig.~11 of~\cite{Abdallah:2002xz}).

The true test of the approach however lies in a verification of the
universality hypothesis, namely that $\alpha_0$ is the same for all
observables and processes.\footnote{Given that one is interested in
  $\alpha_0$, one may wonder why one also fits for $\alpha_s$. The two
  principal reasons are (a) that one is in any case interested in the
  value of $\as$ and (b) that the data and perturbative prediction
  differ also by higher-order terms in $\as$ and fixing $\as$ would
  mean trying to fit these higher-order terms with a power correction,
  which would be misleading.} %
While there are strong general reasons for expecting universality
within $\ee$ (boost invariance and the flux-tube model), universality
across difference processes is less trivial --- for example one could
imagine DIS hadronisation being modified by interactions between
outgoing low-momentum gluons (or hadrons) and the `cloud' of partons
that make up the proton remnant (present even for a fast-moving
proton, since the longitudinal size of the cloud is always of order
$1/\Lambda$).

\begin{figure}[htbp]
  \centering
  \includegraphics[width=0.48\textwidth]{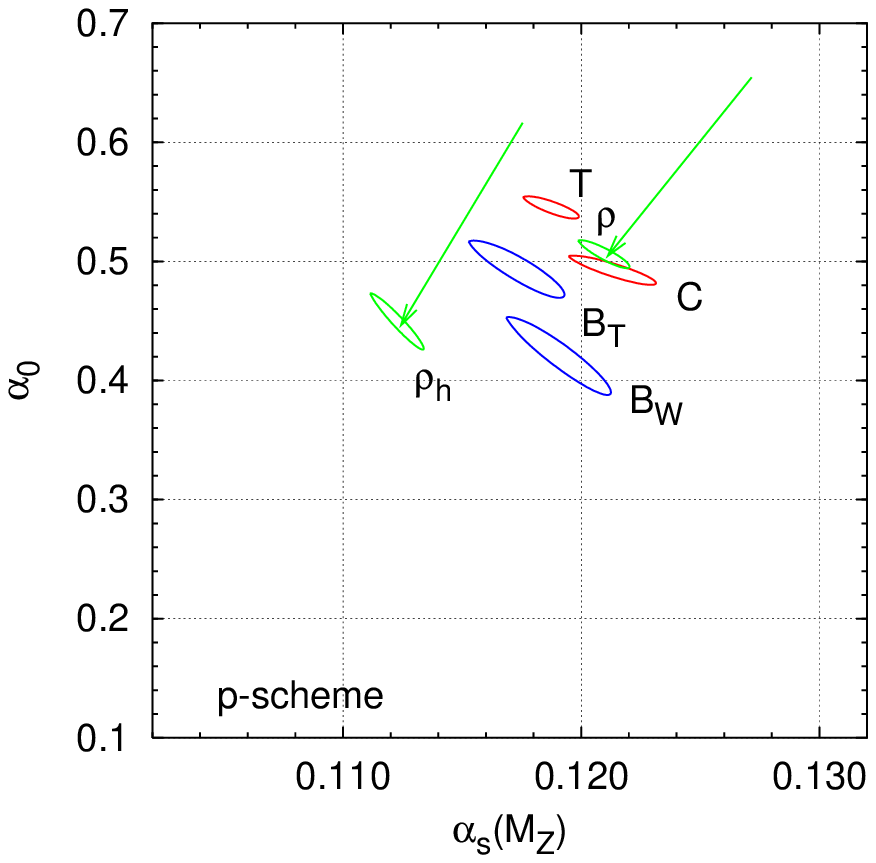}
  \hfill
  \includegraphics[width=0.48\textwidth]{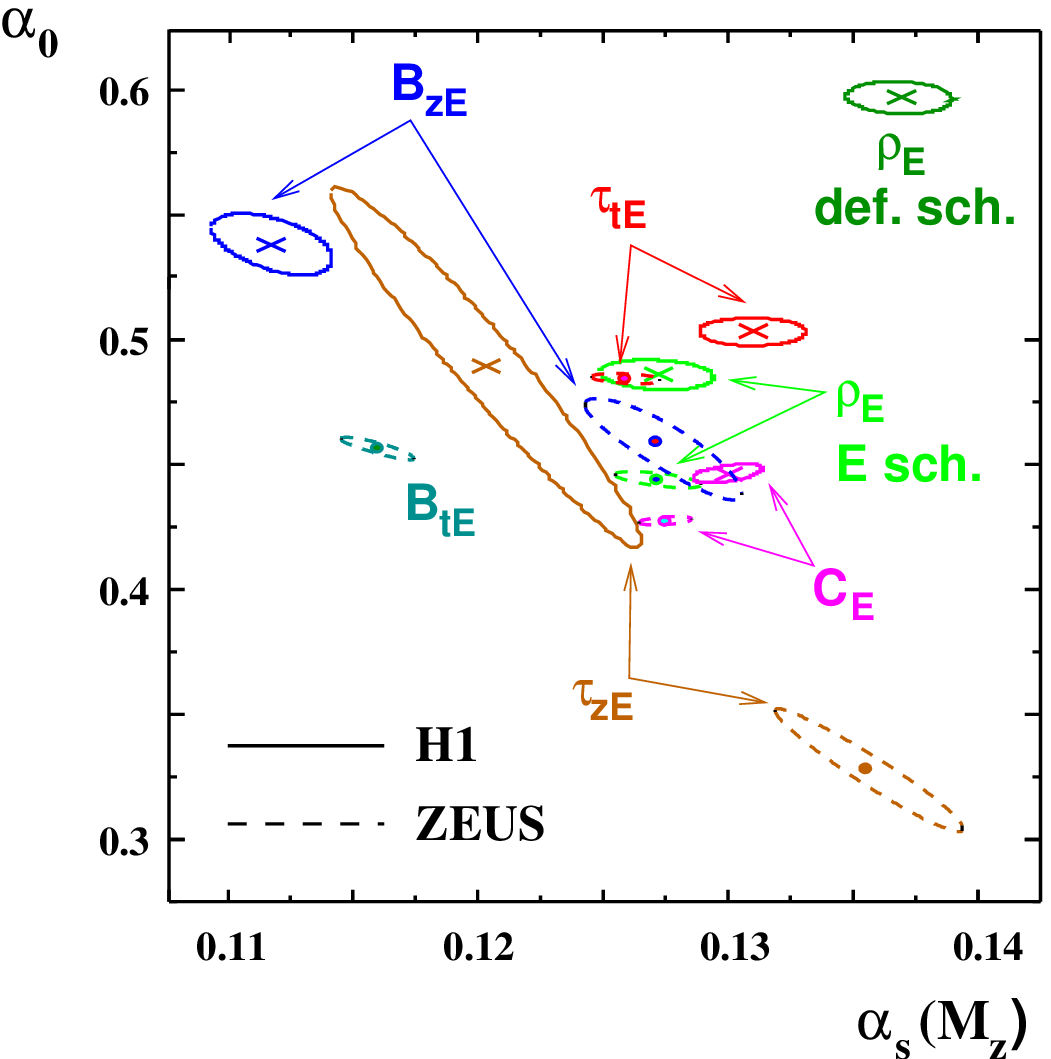}
  \caption{Left: $\as$, $\alpha_0$ fit to $\ee$ data on mean values of
    event shapes, with the jet masses converted to the $p$-scheme (the
    arrow indicates the change in going from default to $p$-scheme)
    \cite{Salwick}. Right: fits to DIS mean event-shape data by H1
    \cite{Adloff:2000,Martyn:2000jk} and ZEUS~\cite{ZEUSmeans}
    (curves taken from figures in~\cite{Martyn:2000jk,ZEUSmeans}). Contours
    indicate 1-$\sigma$ uncertainties (statistical and experimental
    systematic, except for the ZEUS results which are just
    statistical).}
  \label{fig:meanfits}
\end{figure}

Figure~\ref{fig:meanfits} shows $1$-$\sigma$ contours for fits of
$\alpha_s$ and $\alpha_0$ in $\ee$~\cite{Salwick} (left, all experiments
combined) and DIS (right, H1~\cite{Adloff:2000,Martyn:2000jk} and ZEUS
\cite{ZEUSmeans} merged into one plot). Aside from the different
scales used, some care is required in reading the figures because of
different treatments 
of the errors in the different plots. For example H1 include
experimental systematic errors whereas ZEUS do not.\footnote{There are
  differences additionally in the fixed-order predictions used, the H1
  results being based on DISENT~\cite{EVENT2} which was subsequently
  found to have problems~\cite{McCance:jh,DasSalTRC}, whereas the
  final ZEUS results are based on DISASTER++~\cite{Disaster}. The
  differences between fits with DISASTER++ and DISENT are generally of
  similar magnitude and direction~\cite{ZEUSmeans} as those seen
  between H1 and ZEUS results, suggesting that where there is a large
  difference between them (notably for observables measured with
  respect to the photon axis),
  one should perhaps prefer the ZEUS fit result.} %
In the $\ee$ fit, the systematics are included but treated as
uncorrelated, which may underestimate the final error. In neither
figure are theoretical uncertainties included: over the available
range of $\ee$ energies, renormalisation scale dependence gives an
uncertainty of about $\pm 0.005$ on $\as$, while a `canonical'
variation of the Milan factor $\cal M$ by $\pm 20\%$ (to allow for
higher-order terms) leads to an uncertainty of $^{+0.09}_{-0.06}$ on
$\alpha_0$~\cite{MovillaFernandez:2001ed}. In DIS the corresponding
uncertainties are larger for $\as$ ($\pm 0.09)$, essentially because
the fits are dominated by lower $Q$ values, and they vary
substantially for $\alpha_0$ (larger for photon-axis observables,
smaller for the others). Also to be kept in mind is that the DIS
figure includes the jet mass, $\rho_E$ in the default scheme, and in $\ee$
the origins of the arrows indicate the default-scheme results for
$\rho$ and $\rho_H$. Since the default scheme for jet masses breaks
universality (cf.\ section~\ref{sec:DWdetails}) these results should
not be compared directly to those for other observables.

Taking into account the uncertainties, which are not necessarily
correlated from one observable to another, fig.~\ref{fig:meanfits}
indicates remarkable success for the renormalon-inspired picture. The
results are in general consistent with a universal value for
$\alpha_0$ in the range $0.45$ to $0.50$.  Furthermore the $\ee$
results for $\as$ are in good agreement with the world average. On the
other hand the DIS results for $\as$ seem to be somewhat larger than
the world average. The discrepancy is (just) within the theoretical
systematic errors, however it remains a little disturbing and one
wonders whether it could be indicative of large higher-order
corrections, or some other problem (known issues include the fact that
the low-$Q$ data can bias the fits, for example because of heavy-quark
effects and $\Lambda^2/Q^2$ corrections; and also at high-$Q$ there is
$\gamma/Z$ interference which is not usually accounted for in the
fixed-order calculations). There is potentially also some worry about
the $\tau_{zE}$ result which has an anomalously low $\alpha_0$ and
large $\as$ (and additionally shows unexpectedly substantial $x$
dependence~\cite{ZEUSmeans}). 

Another point relates to the default scheme jet masses --- though the
universality-breaking effects should be purely non-perturbative, they
have an effect also on the fit results for $\alpha_s$. This is a
consequence of the anomalous dimension that accompanies $\Lambda/Q$
hadron mass effects. Similar variations of $\as$ are seen when varying
the set of particles that are considered stable~\cite{Salwick}.

Though as discussed above some issues remain, they should not however
be seen as detracting significantly from the overall success of the
approach and the general consistency between $\ee$ and DIS. We also
note that the first moment of the coupling as extracted from studies
of heavy-quark fragmentation \cite{YuriErice,Nason:1996pk} is quite
similar to the value found for event shapes.

There have also been experimental measurements of higher moments of
event shapes, notably in~\cite{Acciarri:2000hm}. The parameters
$\alpha_s$ and $\alpha_0$ are fixed from fits to mean values and then
inserted into eq.~(\ref{eq:momentPC}) to get a prediction for the
second moments. For $\langle\rho_H^2\rangle$ this gives very good
agreement with the data (though the fact that $\rho_H$ is in the
default scheme perhaps complicates the situation), while
$\langle(1-T)^2\rangle$ and $\langle C^2\rangle$ show a need for a
substantial extra correction. In~\cite{Acciarri:2000hm} this extra
contribution is shown to
be compatible with a $\Lambda^2/Q^2$ term, though it would be
interesting to see if it is also compatible with a $\as\cP$ term with
a modified coefficient.


Currently (to our knowledge) no fits have been performed for three-jet
event shapes, though we understand that such fits are in progress
\cite{BanerjeePrivate} for the mean value of the $D$-parameter.  We
look forward eagerly to the results. In the meantime it is possible to
verify standard parameters for $\as$ and $\alpha_0$ against a single
published point at $Q=M_Z$~\cite{Adeva:1992gv,Aleph03} and one finds
reasonable agreement within the uncertainties.

\subsection{Fits with alternative perturbative estimates.}

Two other points of view that have also been used in analysing data on
$e^{+}e^{-}$ event shapes are the approach of {\it{dressed gluons}}
employed by Gardi and Grunberg \cite{GardGrun} and the use of
{\it{renormalisation group improved}} perturbation
theory~\cite{DharGupta} as carried out by the DELPHI
collaboration~\cite{Abdallah:2002xz}.

\subsubsection{Gardi--Grunberg approach.}

In the Gardi-Grunberg approach one uses the basic concept of
renormalons (on which the DW model is also based), but one choses to
treat the renormalon integral differently to the DW model. This
renormalon integral is ill-defined due to the Landau singularity in
the running coupling and instead of assuming an infrared finite
coupling below some matching scale $\mu_I$ as was the case in the DW
model, Gardi and Grunberg define the renormalon integral by its
principal value (a discussion of the theoretical merits of different
approaches has been give in~\cite{Webber:1998um}).  In doing this they
explicitly include higher order 
renormalon contributions in their perturbative result, rather than
including them via power behaved corrections below scale $\mu_I$ that
result from assuming an infrared finite coupling (though they do also
compare to something similar, which they call a `cutoff' approach).
However since one is dealing with an ambiguous integral (a
prescription other than a principal value one would give a result
differing by an amount proportional to a power correction) one must
still allow for a power correction term.  Gardi and Grunberg studied
the mean thrust in $e^{+}e^{-}$ annihilation and used the following
form for fitting to the data :
\begin{equation}
\langle 1-T \rangle = \frac{C_F}{2}[R_{0}|_{\mathrm{PV}} +
\delta_{\mathrm{NLO}}]+\frac{\lambda}{Q} 
\end{equation}
where the subscript PV denotes the principal value of the renormalon
integral $R_0$ for the thrust (which includes the full LO contribution
and parts of the higher-order contributions) and
$\delta_{\mathrm{NLO}}$ is a piece that accounts for the difference
between the true NLO coefficient for $1-T$ and the contribution
included in $R_0$.
\begin{figure}[htbp]
  \centering \includegraphics[width
  =0.55\textwidth,angle=90]{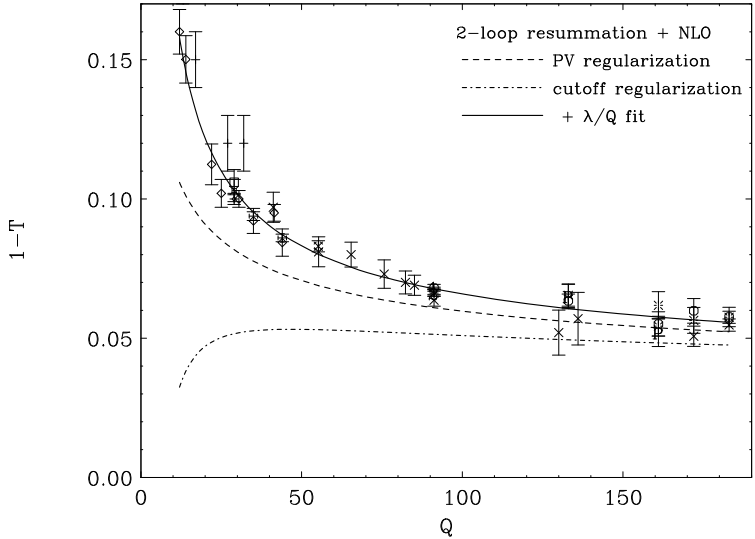}
\caption{The mean value of the thrust distribution in $e^{+}e^{-}$
  annihilation as a function of $Q$ with
  \hbox{$\alpha_s^{\hbox{$\overline{\hbox{\tiny MS}}\,$}}({\rm
      M_Z})=0.110$}. The upper (dashed) line corresponds to the
  principal value result which then is fitted to the data adding a
  $1/Q$ term. Figure taken from Ref.~\cite{Gardtalk}.}
\label{fig:GGthrust}
\end{figure}
The best fit values obtained for
$\alpha_s^{\hbox{$\overline{\hbox{\tiny MS}}$}}({\rm M_Z})$ and
$\lambda$ are respectively $0.110 \pm 0.006$ and $0.62 \pm 0.11$.  We
note that $\lambda\simeq 0.6$ in this approach corresponds to a
smaller power correction than is required in the DW model. This is
probably a consequence of the inclusion of pieces of higher
perturbative orders via a principal value prescription, although it
leads to a somewhat small value (compared to the world average) of the
coupling at scale $M_z$.

\subsubsection{Renormalisation group improved approach.}

Next we turn to the renormalisation group improved (RGI) perturbative
estimates~\cite{DharGupta,Maxwell,Beneke:1993ee} that have also been used to
compare with event shape data for the mean value of different event
shapes in $e^{+}e^{-}$ annihilation.  The basic idea behind this
approach is to consider the dependence of the observable on the scale
$Q$, which can be expressed using renormalisation group invariance as
\begin{equation}
\label{eq:RGI}
Q \frac{dR}{dQ} = -bR^2(1+\rho_1 R+\rho_2 R^2 +\cdots)=b \rho(R)
\end{equation}
where in the above formula $b = \beta_0/2$, $R=2 \langle f \rangle/A$,
with $\langle f \rangle$ being the mean value of a given event shape
and $A$ being the coefficient of $\alpha_s/2\pi$ in its perturbative
expansion.

\begin{figure}[htbp]
\centering
\includegraphics[width=0.4\textwidth]{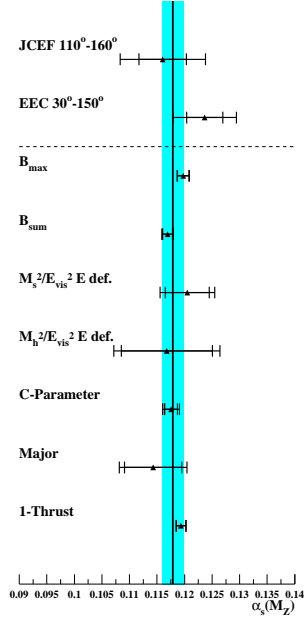}
\caption{Results for $\alpha_s(M_z)$ from comparisons to $e^{+}e^{-}$
  event-shape data using the RGI approach. The band shows the mean
  value of $\as$ and uncertainty, as obtained from the observables
  below the dashed line. Figure taken from
  Ref.~\cite{Abdallah:2002xz}.}
\label{fig:RGI}
\end{figure}

Thus the first term in the perturbative expansion of $R$ is simply
$\alpha_s/\pi$.  The $\rho_i$ are renormalisation scheme independent
quantities. In particular the quantity $\rho_1 = \beta_1/(2 \beta_0)$
is simply the ratio of the first two coefficients of the QCD $\beta$
function while $\rho_2$ additionally depends on the first three (up to
NNLO) perturbative coefficients in the expansion of $\langle f
\rangle$.  Current studies using this method are therefore restricted
to an NLO analysis involving $\rho_1$ alone. With this simplification,
the solution of eq.~(\ref{eq:RGI}) simply corresponds to the
introduction of an observable-specific renormalisation scheme and
associated scale parameter $\Lambda_R$ which is such that it sets the
NLO perturbative term to be zero (and neglects higher perturbative
orders). Its relation to the standard
$\Lambda_{\overline{\mathrm{MS}}}$ is easily obtained:
\begin{equation}
  \frac{\Lambda_R}{\Lambda_{\overline{\mathrm{MS}}}} = e^{\frac{r_1}{b}}
  \left (\frac{2c_1}{b} \right )^{-\frac{c_1}{b}},
\end{equation}
where $c_1 = \rho_1$ and $r_1 = B/2A$, with $B$ being the NLO
coefficient in the $\overline{\mathrm{MS}}$ expansion for the
observable. The terms in the bracket account for different definitions
of the coupling in terms of $\Lambda$ as used in the $R$ and
$\overline{\mathrm{MS}}$ schemes.

In the specific case of fits to $e^{+}e^{-}$ event shape mean
values, the above approach has met with a considerable success
\cite{Abdallah:2002xz}.  In particular the introduction of $\Lambda_R$
seems to remove any need for a significant power correction. The
DELPHI collaboration~\cite{Abdallah:2002xz} have performed a combined
fit of the parameter $\Lambda_R$ and a parameter $K_0$ that quantifies
the power correction~\cite{Maxwell,Abdallah:2002xz}.  The value of
$K_0$ was found to be consistent with zero in most cases and rather
small in all cases which indicates that there is no real need for a
power behaved correction once the perturbative expansion is fixed
through the RGI technique.  The agreement between $\alpha_s$ values
extracted from different observables is impressive (see
figure~\ref{fig:RGI}) with a spread that is only about half as large as
that obtained using the standard perturbative result supplemented by a
power correction term, using the DW initiated model.

While it is clear that the size of the power correction piece
inevitably depends on how one choses to define the perturbative
expansion (in fact all the methods discussed thus far, including the
DW model and the Gardi-Grunberg approach, account for this effect), it
is nevertheless very interesting that for several different
observables, with significantly different perturbative coefficients,
one observes after the introduction of RGI, hardly any need for a
power correction term.  Certainly it is not clear on the basis of any
physical arguments, why the genuine non-perturbative power correction
should be vanishingly small and that a perturbative result defined in
a certain way should lead to a complete description of the data. The
clarification of this issue is still awaited and the above findings
are worth further attention and study.

A final point to be kept in mind about the RGI approach in the above
form, is that it is valid only for inclusive observables that depend
on just a single relevant scale parameter $Q$.  This makes its
applicability somewhat limited and for instance it is not currently
clear how to extend the procedure so as to allow a meaningful study of
event shape distributions in $e^{+}e^{-}$ annihilation or DIS event
shape mean values and distributions (which involve additional scales).


\section{Distributions}
\label{sec:distributions}

\setcounter{footnote}{0}

So far our discussion of comparisons between theory and experiment has
been limited to a study of mean values of the event shapes. As we have
seen, there is some ambiguity in the interpretation of these
comparisons, with a range of different approaches being able to fit
the same data. This is in many respects an unavoidable limitation of
studies of mean values, since the principal characteristic of the
different models that is being tested is their $Q$-dependence, which
can also be influenced by a range of (neglected) higher-order
contributions. In contrast, the full distributions of event shapes
contain considerably more information and therefore have the potential
to be more discriminatory.

In section~\ref{sec:perturbative} we 
discussed the perturbative calculation of distributions. As for
mean values though, the comparison to data is complicated by the need
for non-perturbative corrections. These corrections however involve
many more degrees of freedom than for mean values, and there are
a variety of ways of including them. Accordingly we separate our
discussion into two parts: in section~\ref{sec:evgen} we shall
consider studies in which hadronisation corrections are taken from
Monte Carlo event generators, and where the main object of study is
the perturbative distribution, with for example fits of the strong
coupling. In section~\ref{sec:anldist} we shall then consider studies
which involve analytical models for the non-perturbative corrections,
and where it is as much the non-perturbative models, as the perturbative
calculations that are under study.

\subsection{Perturbative studies with Monte Carlo hadronisation}
\label{sec:evgen}

\begin{figure}[htbp]
  \centering
  \includegraphics[width=0.48\textwidth]{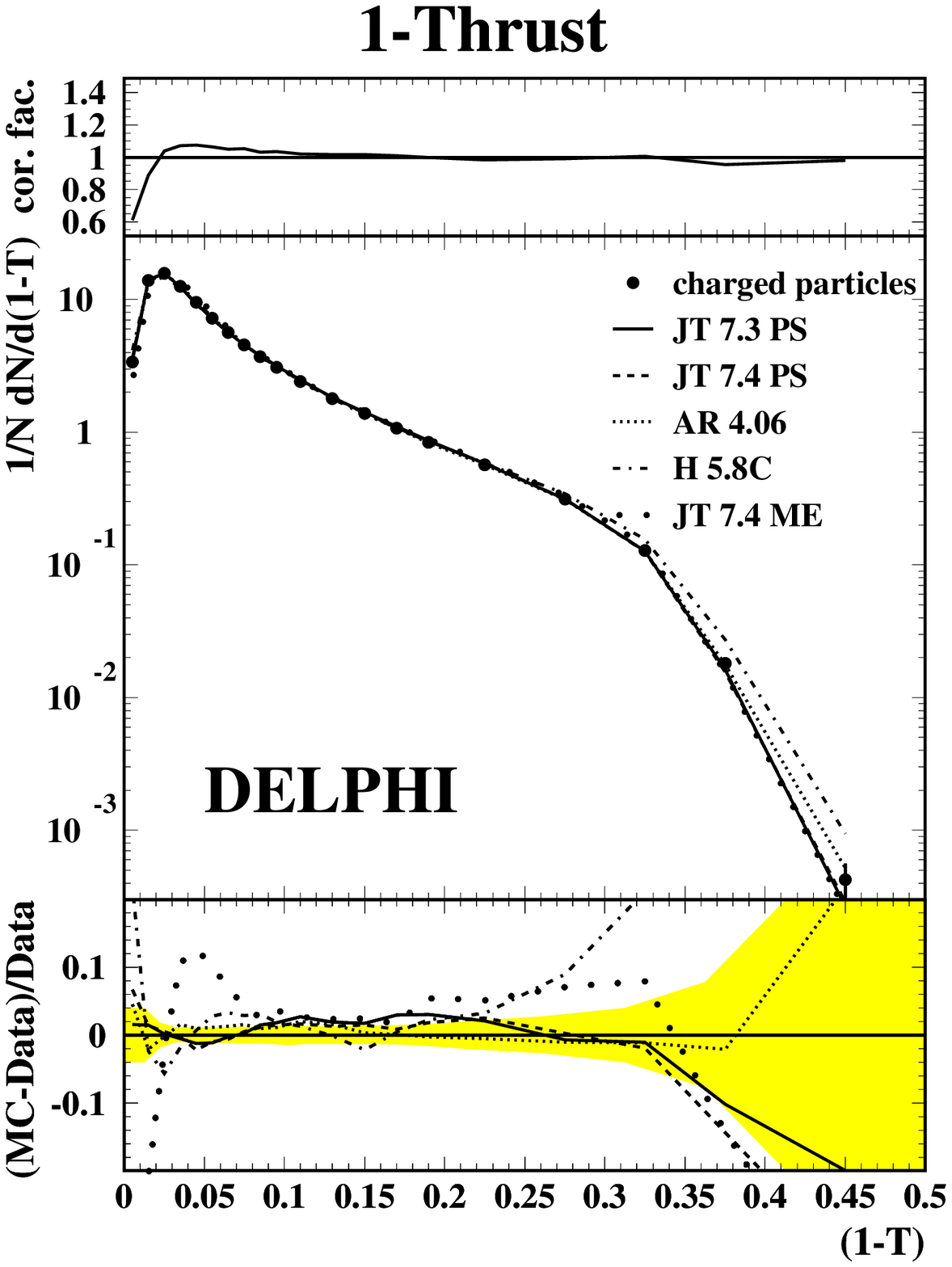}
  \hfill
  \includegraphics[width=0.48\textwidth]{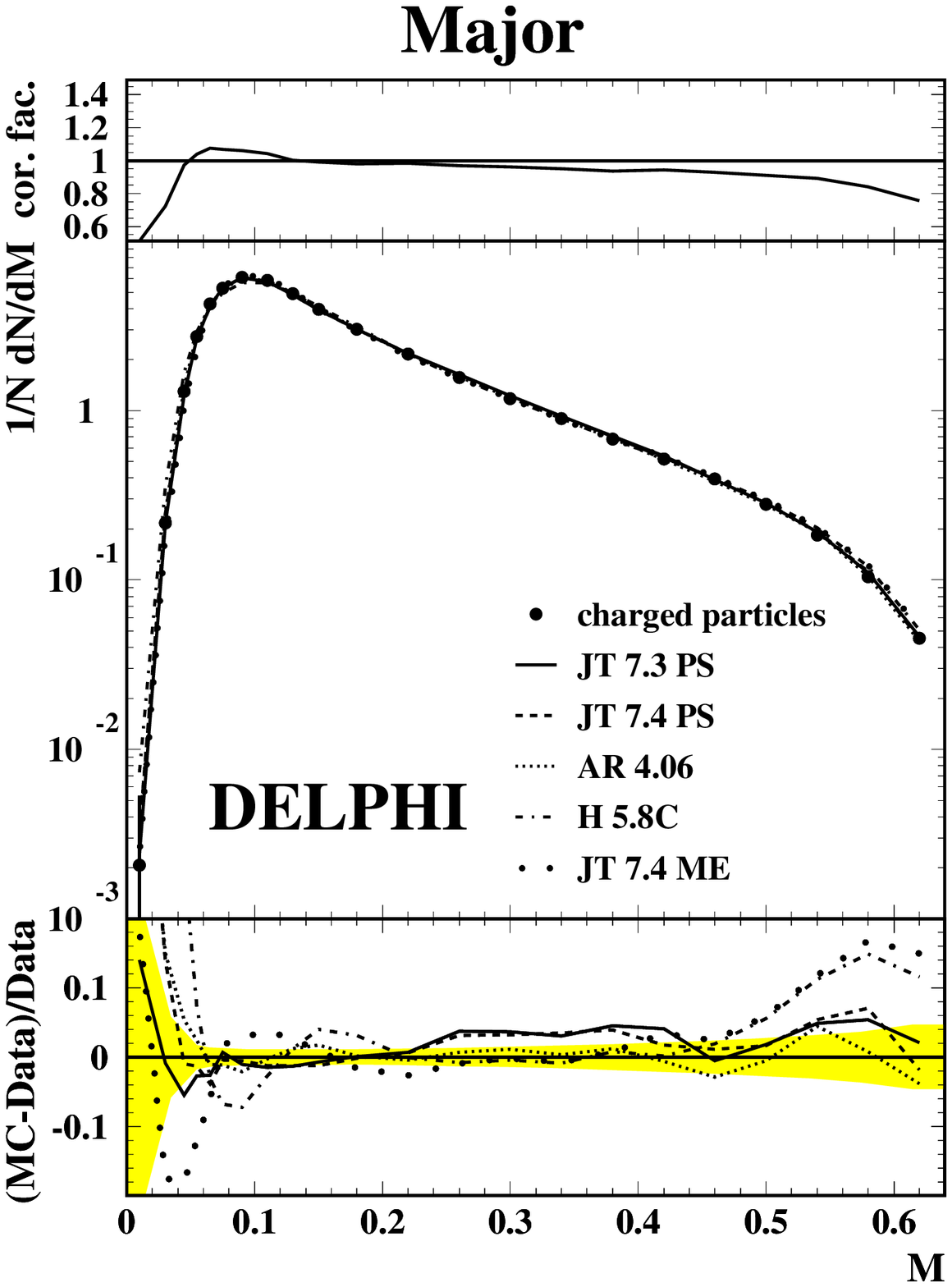}
  \caption{Event shape distributions at $M_Z$ compared
    to a variety of event generators~\cite{Herwig,Pythia,Ariadne}. The
    shaded band in the lower plot indicates the experimental
    statistical and systematic errors added in quadrature. Figure
    taken from~\cite{TuningDelphi}.}
  \label{fig:tuning}
\end{figure}

\subsubsection{Use of Monte Carlo event generators.}
Before considering the purely perturbative studies mentioned above, let
us recall that one major use of event shape distributions has
been in the testing and tuning
\cite{TuningDelphi,TuningAleph,TuningYellowBook} of Monte Carlo event 
generators such as Herwig~\cite{Herwig}, Jetset/Pythia~\cite{Pythia}
and Ariadne~\cite{Ariadne}.  Figure~\ref{fig:tuning} 
shows comparisons for two event shapes, the thrust and
thrust major, and the agreement is remarkable testimony to the ability
of the event generators to reproduce the data.

The use of event generators to probe the details of QCD is
unfortunately rather difficult, essentially because they contain a
number of parameters affecting both the non-perturbative modelling and
in some cases the treatment of the perturbative shower. Furthermore
though considerable progress is being made in matching to fixed order
calculations (see for example~\cite{FrixioneWebber}), event generators
are in general able to guarantee neither the NLL accuracy nor the NLO
accuracy of full matched NLL-NLO resummed calculations.\footnote{It
  is to be noted however that for the most widely-studied $\ee$
  event-shapes (global 
  \cite{NGOneJet} two-jet event shapes) Herwig
  \cite{Herwig,Herwig++} is expected to be correct to NLL accuracy.} %

Nevertheless, the good agreement of the event generators with the data
suggests that the bulk of the dynamics is correct and in particular
that a good model for the hadronisation corrections can be had by
comparing parton and hadron `levels' of the generator. There has been
a very widespread use of event generators in this way to complement
the NLL+NLO perturbative calculations.

Such a method has both advantages and drawbacks and it is worth devoting
some space to them here.
On one hand, the parton
level of an event generator is not a theoretically well-defined
concept --- it is regularised with some effective parton mass or
transverse momentum cutoff, which already embodies some amount of
non-perturbative correction. In contrast the NLO+NLL partonic
prediction integrates down to zero momenta without any form (or need)
of regularisation. This too implies some amount of non-perturbative
contribution, but of a rather different nature from that included via
a cutoff. The resulting difference between the event-generator and the
purely perturbative NLO+NLL parton levels means that the
`hadronisation' that must be applied to correct them to hadron level
is different in the two cases.

\begin{figure}
  \begin{center}
    \includegraphics[width=\textwidth]{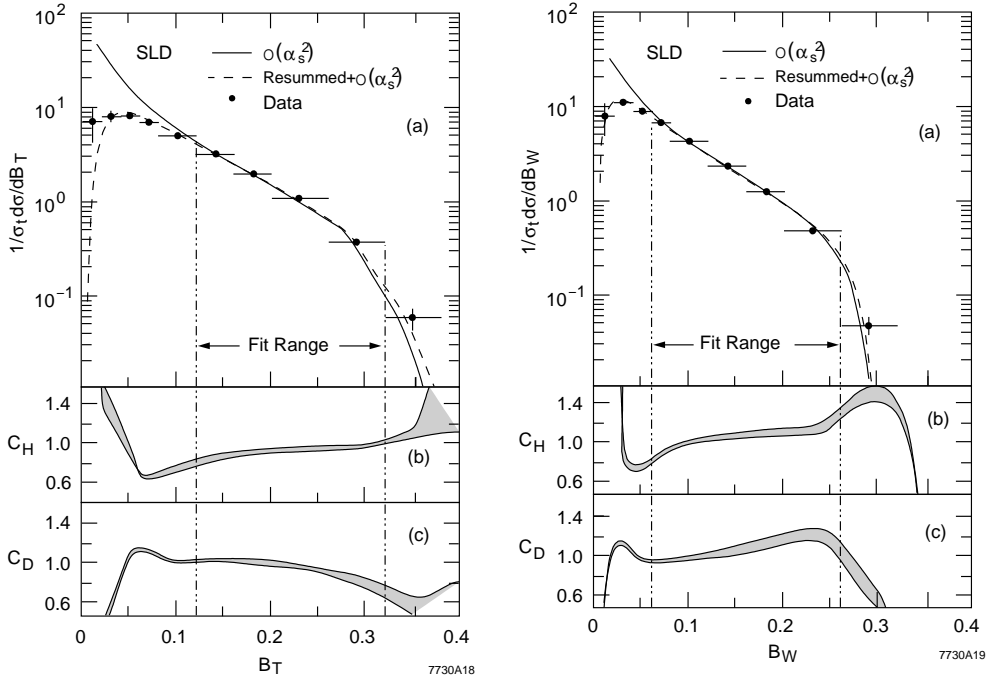}
  \end{center}
  \caption{SLD data for the total and wide-jet broadenings, transformed
    to `parton' level using a multiplicative bin-by-bin factor ($C_H$,
    with uncertainties, as discussed in the text) obtained from the
    ratio of Jetset parton to hadron-level distributions. It is
    compared to NLO and NLL+NLO predictions. Also shown is the
    detector to hadron-level correction factor $C_D$. Figure taken
    from~\cite{Abe:1994mf}.}
  \label{fig:SLDhadr}
\end{figure}

There are nevertheless reasons why event-generators are still used for
determining the hadronisation corrections. The simplest is perhaps
that they give a very good description of the data (cf.\ 
fig.~\ref{fig:tuning}), which suggests that they make a reasonable job
of approximating the underlying dynamics. Furthermore the good
description is obtained with a single set of parameters for all event
shapes, whereas as we shall see, other approaches with a single
(common) parameter are currently able to give equally good
descriptions only for a limited number of event shapes at a time.
Additionally, the objection that the Monte Carlo parton level is
ill-defined can, partially, be addressed by including a systematic
error on the hadronisation: it is possible for example to change the
internal cutoff on the Monte Carlo parton shower and at the same time
retune the hadronisation parameters in such a way that the Monte Carlo
description of the hadron-level data remains reasonable. In this way
one allows for the fact that the connection between Monte Carlo and
NLL+NLO `parton-levels' is not understood.\footnote{A common
  alternative way of determining the Monte Carlo hadronisation
  uncertainty is to examine the differences between the hadronisation
  corrections from different event generators, such as Pythia, Ariadne
  or Herwig. Our theorists' prejudice is that such a procedure is
  likely to underestimate the uncertainties on the hadronisation since
  different event generators are built with fairly similar
  assumptions. On the other hand, it is a procedure that is widely
  used and so does at least have the advantage of being well
  understood.} %
This is the procedure that has been used for example in
\cite{Abe:1994mf}, and the results are illustrated in
figure~\ref{fig:SLDhadr} as a hadronisation correction factor $C_H$
with an uncertainty corresponding to the shaded band.

Two further points are to be made regarding Monte Carlo hadronisation
corrections. Firstly, as is starting now to be well known, it is
strongly recommended not to correct the data to parton-level, but
rather to correct the theoretical perturbative prediction to hadron
level. This is because the correction to parton level entails the
addition of extra assumptions, which a few years later may be very
difficult to deconvolve from the parton-level `data'. It is
straightforward on the other hand to apply a new hadronisation model
to a perturbative calculation.

Secondly it is best not to treat the hadronisation correction as a
simple multiplicative factor. To understand why, one should look again
at figure~\ref{fig:SLDhadr} (which, dating from several years ago,
corrects data to parton level). The multiplicative factor $C_H$
(parton$/$hadron) varies very rapidly close to the peak of the
distribution and goes from above $1$ to below $1$. This is because the
distribution itself varies very rapidly in this region and the effect
of hadronisation is (to a first approximation, see below) to shift the
peak to larger values of the observable. But a multiplicative factor,
rather than shifting the peak, suppresses it in one place and
recreates it (from non-peak-like structure) in another.  Instead of
applying a multiplicative correction, the best way to include a Monte
Carlo hadronisation correction is to determine a transfer matrix
$M_{ij}$ which describes the fraction of events in bin $j$ at parton
level that end up in in bin $i$ at hadron level. Then for a binned
perturbative distribution $P_i$, the binned hadronised distribution
$H_i$ is obtained by matrix multiplication, $H_i = M_{ij} P_i$. This
has been used for example in \cite{Barate:1996fi,Aleph03}.

\subsubsection{NLL+NLO perturbative studies.}
Having considered the basis and methods for including hadronisation
corrections from event generators, let us now examine some of the
perturbative studies made possible by this approach.  The majority of
them use NLL+NLO perturbative calculations to fit for the strong
coupling (as discussed below some, e.g.\ \cite{Abreu:2000ck}, just use
NLO results, which exist for a wider range of observables) and such
fits can be said to be one of the principal results from event-shape
studies. In general there is rather good agreement between the data
and the resummed predictions (cf.\ fig.~\ref{fig:SLDhadr}).

\begin{figure}[htbp]
  \centering
  \includegraphics[width=0.6\textwidth]{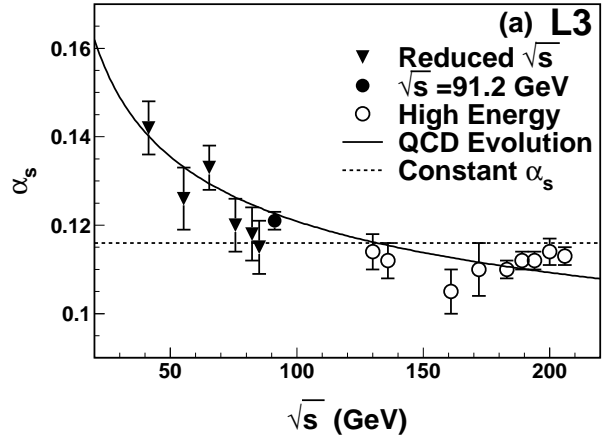}
  \caption{Values for $\as$ obtained by fitting NLL+NLO event-shape
    distributions to L3 data at a range of centre of mass energies
    (figure taken from~\cite{Achard:2002kv}). Only experimental
    uncertainties are shown.}
  \label{fig:L3asEvln}
\end{figure}

An example of such a study is given in figure~\ref{fig:L3asEvln},
which shows results for fits of $\alpha_s$ to L3 data at a range of
energies (including points below $M_Z$ from radiative events which, as
discussed earlier, are to be taken with some caution). One important
result from such studies is an overall value for $\alpha_s$, which in
this case is $\as(M_Z) = 0.1227 \pm 0.0012 \pm 0.0058$
\cite{Achard:2002kv}, where the first error is experimental and the
second is theoretical. Similar results have been given by other
collaborations, e.g.\ $\as(M_Z) = 0.1195 \pm 0.0008 \pm 0.0034$ from
ALEPH~\cite{AlephConf2000027}. There is also good evidence for the
evolution of the coupling, though leaving out the potentially doubtful
radiative points leaves the situation somewhat less clearcut, due to
the limited statistics at high energies (another resummed study which
gives good evidence for the running of the coupling uses JADE and OPAL
data~\cite{Pfeifenschneider:1999rz}, however it is limited to jet
rates).

These and similar results from the other LEP
collaborations and JADE highlight two important points. Firstly at higher
energies there is a need to combine results from all the LEP
experiments in order to reduce the statistical errors and so improve
the evidence for the running of the coupling in the region above
$Q=M_Z$. Work in this direction is currently in progress
\cite{LEPQCDComb}. So far only preliminary results are available
\cite{Stenzel:2003qx}, and as well as improving the picture of the
high-energy running of the coupling, they suggest a slightly smaller
value of $\alpha_s$ than that quoted above, more in accord with world
averages~\cite{Bethke:2002rv,Hagiwara:fs}.

\begin{figure}[htbp]
\centering
\includegraphics[width=0.63\textwidth]{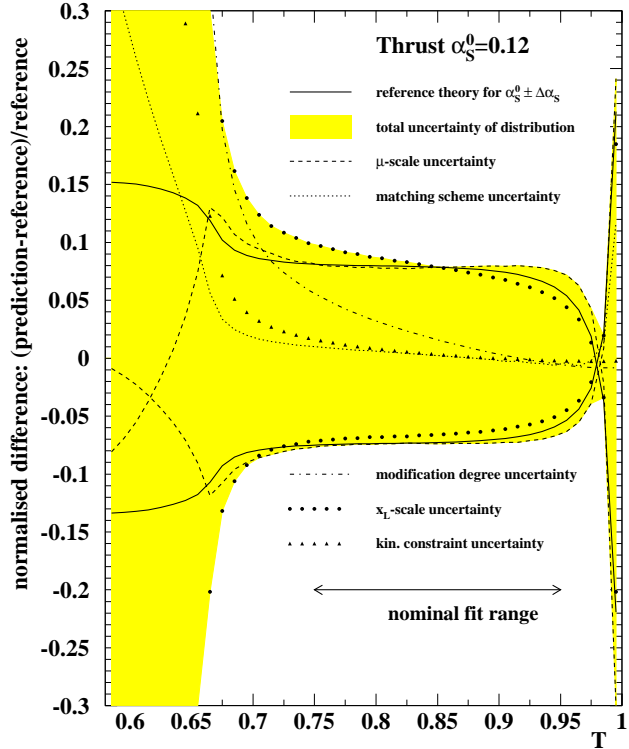}\hfill
\caption{Theoretical uncertainties for the thrust. The shaded area
  represents the overall theoretical uncertainty on the distribution,
  determined as the maximum/minimum of all the dashed, dotted, etc.\ 
  curves; the solid curves indicate the variation of $\alpha_s$
  allowed such that the prediction stays within the uncertainty band.}
  \label{fig:LEPQCDbands}
\end{figure}

Secondly, in the overall result for $\as(M_Z)$, by far the most
important contribution to the error is that from theoretical
uncertainties. However theoretical uncertainties are notoriously
difficult to estimate, since they relate to unknown higher orders.
Systematic investigations of the various sources of uncertainty have
been carried out in~\cite{DasSalTRC,LEPQCDbands} and in particular
\cite{LEPQCDbands} proposes a standard for the set of sources of
uncertainty that ought to be considered, together with an approach for
combining the different sources into a single overall uncertainty on
$\as$.  One of the main principles behind the method is that while one
may have numerous estimates of sources of theoretical uncertainty, it
is not advisable to combine them in quadrature, as one might be
tempted to do, because this is likely to lead to a double-counting of
uncertainties. Rather, one should examine the different sources of
uncertainty across the whole range of the distribution and at each
point, take the maximum of all sources to build up an uncertainty
envelope or band, represented by the shaded area in
figure~\ref{fig:LEPQCDbands} (shown relative to a reference prediction
for a `standard' theory). The overall uncertainty on the coupling
(rather than the distribution) is given by the range of variation of
$\as$ such that the prediction remains within the band. It is to be
noted that this kind of approach is of relevance not just to
event-shape studies but also quite generally to any resummed matched
calculation, which is inherently subject to many sources of
arbitrariness.

\subsubsection{Other perturbative studies.}
One of the drawbacks of NLL+NLO studies is that the NLL resummed
predictions exist for only a fraction of observables.  Recently there
has been an extensive study by the DELPHI collaboration
\cite{Abreu:2000ck} of 18 event-shape distributions compared to NLO
calculations. Since in most cases NLO calculations with a
renormalisation scale $\mu_R=Q$ tend to describe the data rather
poorly, they have examined various options for estimating higher
orders so as to improve the agreement. Their main approach involves a
simultaneous fit of the coupling and the renormalisation scale, i.e.\ 
the renormalisation scale dependence is taken as a way of
parameterising possible higher orders.  In such a procedure the choice
of fit-range for the observable is rather critical, since the true
higher-orders may involve have quite a different structure from that
parameterised by the $\mu_R$ dependence. The DELPHI analysis restricts
the fit range to be that in which a good fit can be obtained. With
these conditions they find relatively consistent values of $\as$
across their whole range of observables (finding a scatter of a few
percent), with a relatively modest theoretical uncertainty, as
estimated from the further variation of $\mu_R$ by a factor of $1/2$
to $2$ around the optimal scale. Their overall result from all
observables treated in this way is $\as = 0.1174\pm 0.0026$ (including
both experimental and theoretical errors).

They also compare this approach to a range of other ways of
`estimating' higher orders. They find for example that some
theoretically based methods for setting the scale (principle of
minimal sensitivity~\cite{Stevenson:1981vj}, effective charge
\cite{Grunberg:1982fw}) lead to scales that are quite correlated to
the experimentally fitted ones, while another method (BLM
\cite{Brodsky:1982gc}) is rather uncorrelated. They compare also to
resummation approaches, though there is no equivalent
renormalisation scale choice that can be made. Instead the comparison
involves rather the quality of fit to the distribution and the final
spread of final $\alpha_s$ values as a measure of neglected
higher-order uncertainties. Not surprisingly, they find that
while the NLL results work well in the two-jet region, in the
three-jet region the combined NLL+NLO predictions fare not that much
better than the pure ($\mu_R=Q$) NLO results. They go on to argue that
NLL+NLO is somewhat disfavoured compared to NLO with an `optimal'
scale.  This statement is however to be treated with some caution,
since the fit range was chosen specifically so as to obtain a good fit
for NLO with the `optimal' scale --- one could equally have chosen a
fit range tuned to the NLL+NLO prediction and one wonders whether the
NLO with optimal scale choice would then have fared so well.
Regardless of this issue of fit range, it is interesting to note that
over the set of observables which can be treated by both methods, the
total spread in $\alpha_s$ values is quite similar, being $\pm 0.04$
in the case of NLO with optimal scale choice and $\pm 0.05$ for
NLL+NLO. This is similar to the spread seen for fits to mean values
with analytical hadronisation models (fig.~\ref{fig:meanfits}), though
the RGI fits of~\cite{Abdallah:2002xz} have a somewhat smaller error.

The jet rates (especially the Durham and Cambridge algorithms and
certain members of the JADE family) are among the few observables for
which the pure NLO calculation gives a reasonable description of the
distribution (cf.\ table~3 of~\cite{Abreu:2000ck}).  One particularly
interesting set of NLO studies makes use of the 3-jet rate as a
function of $y_{cut}$ (this is just the integral of the distribution
of $y_{23}$) in events with primary $b$-quarks as compared to
light-quark events 
\cite{Abreu:1997ey,Abbiendi:2001tw,Brandenburg:1999nb,Barate:2000ab,DelphiConfrunningb}
(some of the analyses use other observables, such as the 4-jet rate or
the thrust).  Using NLO calculations which account for massive quarks
\cite{Bernreuther:1997jn,Rodrigo:1997gy,Nason:1997nw} makes it
possible, in such studies,
to extract a value for the $b$ mass at a renormalisation scale of
$M_Z$, giving first evidence of the (expected) running of the $b$-quark
mass, since all the analyses find $M_b(M_Z)$ in the range $2.6$ to
$3.3$~GeV (with rather variable estimates of the theoretical error).


%
%
%
%
%
%



\subsection{Studies with analytical hadronisation models}
\label{sec:anldist}

We have already seen, in section~\ref{sec:meanPC}, that analytical
models for hadronisation corrections, combined with normal
perturbative predictions, can give a rather good description of mean
values. There the inclusion of a non-perturbative (hadronisation)
correction was a rather simple affair, eq.~(\ref{eq:meancorrections}),
since it was an additive 
procedure. In contrast the general relation between the full
distribution $D_{V}(v)$ for an observable $V$ and the perturbatively
calculated distribution $D_{\mathrm{PT},V}$ is more complicated,
\begin{equation}
  \label{eq:shapeGenConvoluion}
  D_{V}(v) = \int dx \,f_V(x, v, \as(Q), Q)\,
  D_{\mathrm{PT},V}\left(v - \frac{x}{Q}\right)\,,
\end{equation}
where $f_V(x, v, \as(Q), Q)$ encodes all the non-perturbative
information. Eq.~(\ref{eq:shapeGenConvoluion})
can be seen as a generalised convolution, where the shape of the
convolution function depends on the value $v$ of the variable as well
as the coupling and hard scale. Such a general form contains far more
information however than can currently be predicted, or even
conveniently parameterised. Accordingly analytical hadronisation
approaches generally make a number of simplifying assumptions,
intended to be valid for some restricted range of the observable.

\subsubsection{Power correction shift.}
The most radical simplification of eq.~(\ref{eq:shapeGenConvoluion})
that can be made is to replace $f_V(x,v,\as(Q),Q)$ with a
$\delta$-function, $\delta(x - C_V \mathcal{P})$, leading to
\begin{equation}
  \label{eq:NPshift}
  D_{{\mathrm{NP}},V}(v) = D_{{\mathrm{PT},V}}(v-C_V \mathcal{P})\,,
\end{equation}
This was proposed and investigated in~\cite{dok3},\footnote{Related
  discussions had been given earlier~\cite{korch1,korch2}, but the
  approach had not been pursued in detail at the time.} %
and the combination $C_V\mathcal{P}$ is the same that appears in the
power correction for the mean value~\cite{dok1}, discussed in
section~\ref{sec:meanPC}.  Eq.~(\ref{eq:NPshift}) holds for
observables with an $\alpha_s$ independent power correction in
tables~\ref{tab:cvs} and \ref{tab:cvsDIS}, in the region $\Lambda/Q
\ll v \ll 1$, where the lower limit ensures that one can neglect the
width of the convolution function $f_V$, while the upper limit is the
restriction that one be in the Born limit (in which the $C_V$ were
originally calculated). For certain 3-jet observables a similar
picture holds, but with a shift that depends on the kinematics of the
3-jet configuration~\cite{BDMZtmin,BDMZdpar}.

For the broadenings, the situation is more complicated because the
power correction is enhanced by the rapidity over which the quark and
thrust axes can be considered to coincide, $\ln 1/\theta_{Tq}$,
$\theta_{Tq}$ being the angle between thrust and quark axes --- this
angle is strongly correlated with the value of the broadening
(determined by perturbative radiation) and accordingly the extent of
the shift becomes $B$-dependent~\cite{mil3}. The simplest case is the
$\ee$ wide-jet broadening, for which one has
\begin{equation}
  \label{eq:NPbroadshift}
  \Sigma_{\mathrm{NP},B_W}(v) = \Sigma_{\mathrm{PT},B_W}\left(v -
    \frac{D_1(v)\mathcal{P}}{2} \right),\qquad D_1(v) \sim \ln \frac1v \,,
\end{equation}
where the shift is written for the integrated distribution
$\Sigma_{\mathrm{NP},B_W}(v)$ in order to simplify the expressions.
The $1/\sqrt{\as}$ enhancement for the power correction to the mean
broadening (tables~\ref{tab:cvs} and \ref{tab:cvsDIS}) comes about
simply because the average value of $\ln \frac1v$, after integration
over $v$ with the resummed distribution, is of order $1\sqrt{\as}$.

The full form for $D_1(v)$ and analogous results for the total $\ee$
and DIS broadenings have been given in~\cite{mil3,DasSalBroad}, with
results existing also for the thrust minor~\cite{BDMZtmin}, and the
DIS and Drell-Yan out-of plane momenta~\cite{KOUTDIS,KOUTDY}. Yet
subtler instances of perturbative, non-perturbative interplay arise
for observables like the EEC~\cite{EECresum} and DIS azimuthal
correlation~\cite{AZIMDIS}, with the appearance of \emph{fractional}
powers of $Q$ in the power correction, as mentioned
before.

\begin{figure}[htbp]
  \centering
  \includegraphics[width=0.44\textwidth,height=0.48\textwidth,angle=90]{btdist-opal91.eps}
  \hfill
  \includegraphics[width=0.48\textwidth]{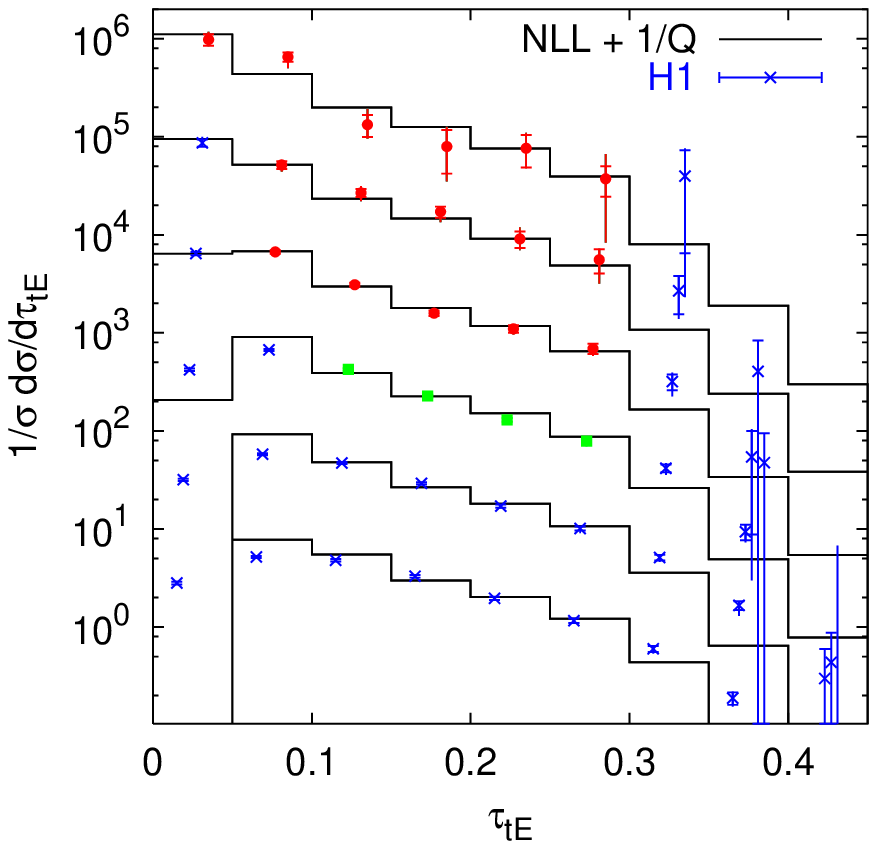}
  \caption{Left: Illustration of the $B_T$-dependent non-perturbative shift in
    the jet broadening distribution for $e^{+}e^{-}$ annihilation. The
    dashed curve is the perturbative prediction (NLL+NLO), while the
    solid curve includes a non-perturbative ($B$-dependent) shift.
    Figure taken from~\cite{mil3}. Right: NLL+NLO distribution with a
    $1/Q$ shift for $\tau_{tE}=1-T_{tE}$ in DIS at a range of $Q$ values (round
    dots indicate points used in fits)~\cite{DasSalTRC}. The $Q$
    values range from 15~GeV (bottom) to 81~GeV (top).}
  \label{fig:broadpc}
\end{figure}
One might a priori have thought that the formal domain of validity of
the shift approximation, $\Lambda/Q \ll v \ll 1$, would be somewhat
limited.  Figure~\ref{fig:broadpc} (left) illustrates what happens in
the case of $B_T$ --- quite remarkably the shift describes the data
well over a large range of $B_T$, with slight problems only in the
extreme two-jet region (peak of the distribution) and in the four-jet
region ($B_T \gtrsim 0.3$). Similar features are seen in the
right-hand plot of fig.~\ref{fig:broadpc} for $\tau_{tE}$ in DIS at a
range of $Q$ values: a large part of the distribution is described
for all $Q$ values, and problems appear only in the $3+1$ jet region
($\tau_{tE}\gtrsim 0.3$), and in the $1+1$ jet region for low $Q$.
This success is reproduced for quite a range of observables in $\ee$
and DIS.

\begin{figure}[t]
  \centering
  \includegraphics[width=0.48\textwidth]{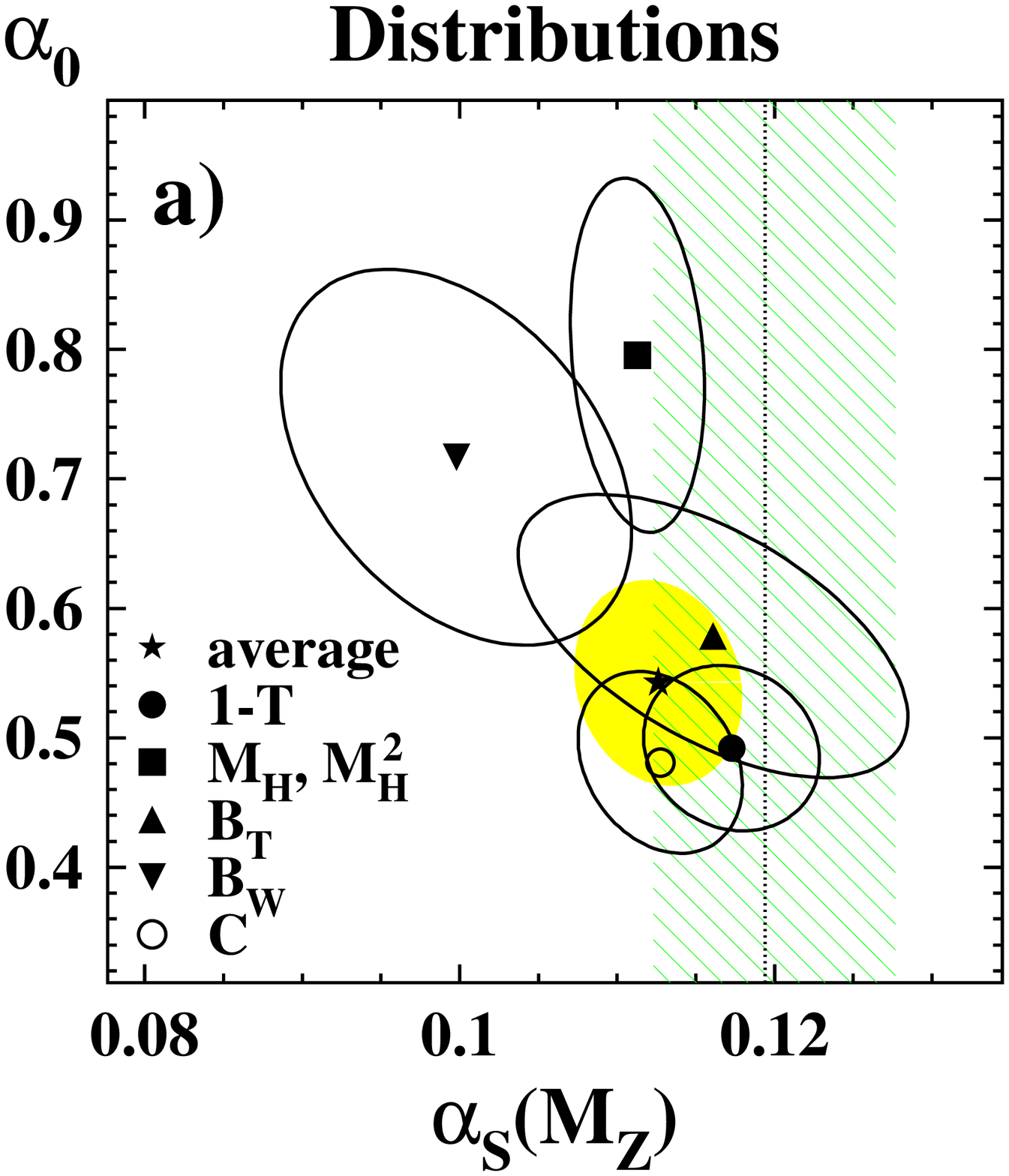}\hfill
  \includegraphics[width=0.48\textwidth]{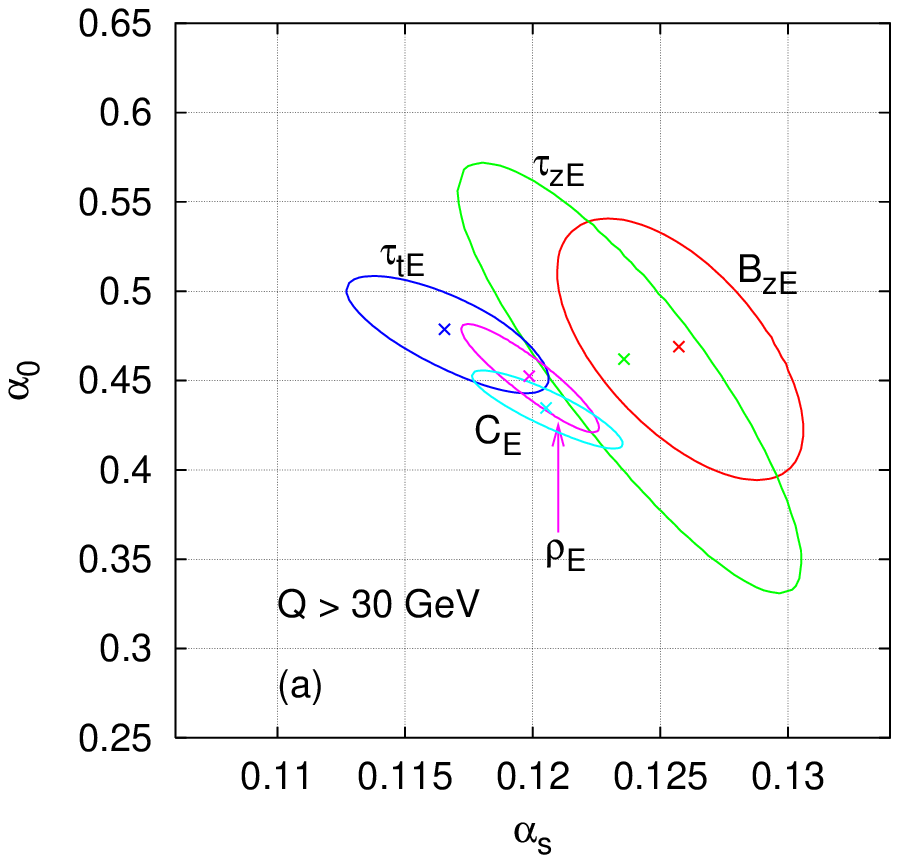}
  \hfill
  \caption{Left: $1$-$\sigma$ contours for simultaneous fits of $\as$
    and $\alpha_0$ to $\ee$ event-shape distributions. The error
    contours account for statistic, systematical and theoretical
    uncertainties. The shaded band indicates the result for $\alpha_s$
    when using Monte Carlo hadronisation models.  Figure taken from
    \cite{MovillaFernandez:2002hu}.  Right: $1$-$\sigma$ contours from
    fits to H1 DIS event shape distributions~\cite{Adloff:2000} with
    statistical and systematic errors added in quadrature. Figure
    taken from~\cite{DasSalTRC}.}
  \label{fig:distcont}
\end{figure}
As was the case for mean values, a more systematic test involves
carrying out a simultaneous fit of $\as$ and $\alpha_0$ for each
observable and then checking for consistency between observables and
with the corresponding results for mean values. Results from $\ee$ and
DIS are shown in figure~\ref{fig:distcont}, taken from
\cite{MovillaFernandez:2001ed,MovillaFernandez:2002hu} and
\cite{DasSalTRC}. Other fits that have been carried out in recent
years
include~\cite{AlephConf2000027,Abdallah:2002xz,Kluge:2003sa}.\footnote{While
  this manuscript was being completed, new analyses by ALEPH
  \cite{Aleph03} were made public; these are not taken into account in
  our discussion, though the picture that arises is similar to that
  in \cite{MovillaFernandez:2002hu}. We note also that \cite{Aleph03}
  provides the first (to our knowledge) publicly available data on the
  thrust minor and $D$-parameter in $3$-jet events. }

Just as for the mean values, one notes that (with exception of $B_W$
and $M_H^2$ in $\ee$, to be discussed shortly) there is good
consistency between observables, both for $\ee$ and DIS. This
statement holds holds also for the EEC, fitted in
\cite{Abdallah:2002xz}, not shown in fig.~\ref{fig:distcont}.  There
is also good agreement with the results for the mean values,
fig.~\ref{fig:meanfits}, except marginally for DIS $\as$ results,
which for distributions are in better accord with the world average.

The good agreement between
distributions and mean values is important in the light of alternative
approaches to fitting mean values, such as the RGI which, we recall,
seems to suggest that the power correction
for mean values can just as well be interpreted as higher order
contributions.  Were this really the case then one would expect to
see no relation between $\alpha_0$ results for mean values and
distributions.\footnote{%
  We recall that while the RGI approach for mean values shows no need
  for a power correction~\cite{Maxwell,Abdallah:2002xz}, the optimal
  renormalisation scale approach~\cite{Abreu:2000ck} for
  distributions, which shows a strong correlation to the effective
  charge approach (which itself is equivalent to the RGI approach),
  has to explicitly incorporate a Monte Carlo hadronisation
  correction.}

Despite the generally good agreement, some problems do persist.
Examining different groups' results for the $\alpha_s$-$\alpha_0$ fits
one finds that for some observables there is a large spread in the
results. For example for the $\ee$ $B_T$, ALEPH
\cite{AlephConf2000027} find $\alpha_s = 0.109 \pm 0.007$ while the
JADE result~\cite{MovillaFernandez:2002hu} shown in
fig.~\ref{fig:distcont} corresponds to $\alpha_s = 0.116 \pm 0.012$
(and this differs from a slightly earlier JADE result $\alpha_s =
0.111 \pm 0.006$~\cite{MovillaFernandez:2001ed}). While the results
all agree to within errors, those errors are largely theoretical and
would be expected to be a common systematic on all results. In other
words we would expect the results to be much closer together than
$1$-$\sigma$. That this is not the case suggests that the fits might
be unstable with respect to small changes in details of the fit, for
example the fit range (as has been observed when including data in the
range $20 <Q < 30$ in the fit corresponding to the right hand plot of
figure~\ref{fig:distcont}~\cite{DasSalTRC}). These differences can
lead to contradictory conclusions regarding the success of the
description of different observables and it would be of value for the
different groups to work together to understand the origin of the
differences, perhaps in a context such as the LEP QCD working group.

There do remain two cases however where there seems to be clear
incompatibility between the data and theoretical prediction: the
heavy-jet mass, $\rho_H$ (also referred to as $M_H^2$) and the
wide-jet broadening, $B_W$. The heavy-jet mass is
subject to $\Lambda/Q$ hadron mass effects, as discussed already in
the context of mean values, however even after
accounting for this by considering $p$ or $E$-scheme data, as has been
done by DELPHI~\cite{Abdallah:2002xz}, the results appear to be rather
inconsistent with other observables: $\alpha_s$ is too small and
$\alpha_0$ too large, as for $B_W$,\footnote{We note 
that fits~\cite{Salam:1999kx} for $B_W$ with a fixed $\as = 0.118$
lead to a consistent value for $\alpha_0 \simeq 0.50$,
though the overall fit quality is poor.} though less extreme. The
common point between these two observables is that they both select
the heavier/wider hemisphere and so are less inclusive than other
observables. Studies with event generators~\cite{Salwick} suggest that
whereas for more inclusive observables the 2-jet limit of the shift
works well even into the 3-jet region, for heavy-wide observables the
3-jet limit behaves very differently from the 2-jet limit and these
differences becomes relevant even for relatively low values of the
observable (some partially related discussion of the problem has been
given also in~\cite{GRmass}). We will come back to this and other
potential explanations below when we discuss shape functions.

Beyond the `usual' observables discussed above, there are some recent
results also for the $C$-parameter shoulder at $C=3/4$, where
\cite{Abdallah:2002xz} the shoulder position is seen to be consistent
with a shift with the usual 2-jet coefficient $C_C=3\pi$ --- this is
in contradiction with the (smaller) result eq.~(\ref{eq:C34power}) for
the shift to the shoulder; however the analysis~\cite{Abdallah:2002xz}
doesn't take into account resummation effects at the shoulder which
may also contribute to an effective shift. A first analysis of
corrections to the $y_3$ distribution has been given in
\cite{Kluth:2003uq} which shows evidence for $1/Q^2$ contributions in
the $y_3$ distribution at the two lowest JADE centre of mass energies
(14 and 22~GeV).

\begin{figure}[htbp]
  \centering \includegraphics[width=0.51\textwidth]{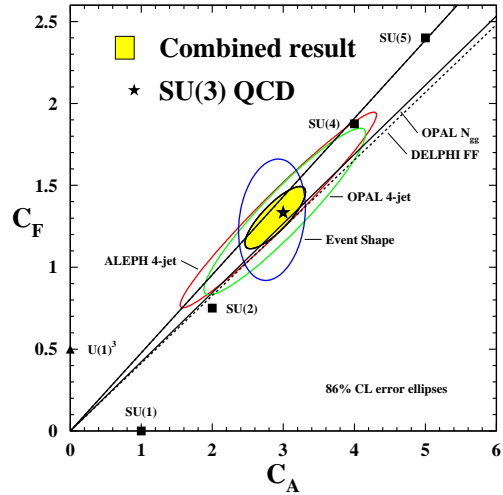}
  \caption{Determination of colour factors from resummed
    distributions with $1/Q$ (shift) power corrections and a variety
    of other methods. Figure taken from~\cite{Kluth:2003yz}. The
    ellipses show correlated measurements from studies of the angular
    distributions of 4-jet events
    \cite{Heister:2002tq,Abbiendi:2001qn} and event shapes
    \cite{Kluth:2000km}, while the lines are obtained from studies of
    the gluon fragmentation function~\cite{Abreu:1999af} and from
    multiplicities in gluon jets~\cite{Abbiendi:2001us}.}
  \label{fig:colfact}
\end{figure}

One last interesting point to discuss regarding the shift
approximation for hadronisation is an application to the measurement
of the QCD colour factors. The fact that the hadronisation is encoded
via just a single parameter, with all the dependence on colour factors
available explicitly for the perturbative and non-perturbative
contributions, means that it is feasible to carry out a simultaneous
fit not only for $\alpha_s $ and $\alpha_0$ but also for one or more
of the QCD colour factors $C_A$, $C_F$ and $T_R$~\cite{Kluth:2000km}.
This would have been much more difficult using Monte Carlo
hadronisation because all the hadronisation parameters would have had
to have been retuned for each change of the colour factors. The
results for this fit are summarised and compared to other approaches
in figure~\ref{fig:colfact}~\cite{Kluth:2003yz}, which shows
$2$-$\sigma$ contours for $C_F$ and $C_A$. While other approaches
constrain the ratio $C_A/C_F$, the event-shape approach gives by
far the best independent measurement of $C_A$ and $C_F$. As a result
the combination of the results gives rather tight limits on the colour
factors, which are in good agreement with the QCD expectation.

\subsubsection{Shape functions.}

We have already mentioned that the approximation of a simple shift is
valid only in the region $\Lambda/Q \ll v \ll 1$. To better understand the
origin of the lower limit, let us expand eq.~(\ref{eq:NPshift}),
\begin{equation}
  \label{eq:expandedshift}
  D_{\mathrm{NP},V}(v) =   D_{\mathrm{PT},V}\left(v\right)
 + \sum_{n=1}^\infty (-1)^n \frac{(C_V {\cal P})^n}{n!}
 D_{\mathrm{PT},V}^{(n)}\left(v\right)
\end{equation}
where $D_{\mathrm{PT},V}^{(n)}(v)$ is the $n^\mathrm{th}$ derivative
of $D_{\mathrm{PT},V}(v)$ with respect to $v$. Since
$D_{\mathrm{PT},V}^{(n)}(v)$ goes as $1/v^n$ one sees that when the
shift $v\sim C_V \cP$ all terms in the series are of the same order.
While the $n=1$ term can be related to the power correction to
the mean value of $V$, the shift merely provides an ansatz for the
higher terms. The breakdown for small values of $vQ$ is clearly
visible for the (DIS) thrust in fig.~\ref{fig:broadpc} (right).

A physically transparent way of dealing with the $n\ge 2$ terms has
been developed by Korchemsky, Sterman and collaborators
\cite{korch1,korch2,Korchemsky:1998ev,korch3,korch4,korch5}. One
approximates $f_V(x,v,\as,Q)$ in eq.~(\ref{eq:shapeGenConvoluion}) by
a \emph{shape function}~\cite{ShapeFunctions} $\tilde f_V(x)$ (rather
than the $\delta$-function which corresponds to a pure shift),
\begin{equation}
  \label{eq:shapefunction}
  D_{\mathrm{NP},V}(v) = \int dx {\tilde f}_V(x)\, D_{\mathrm{PT},V}
  \left(v-\frac{x}{Q}\right).
\end{equation}
In eq.~(\ref{eq:expandedshift}), $(C_V P)^n$ is then replaced by
${\tilde f}_n/Q^n$ where ${\tilde f}_n$ is the $n^\mathrm{th}$ moment
of ${\tilde f}_V(x)$. 

\begin{figure}[htbp]
  \centering
  \includegraphics[width=0.48\textwidth,height=0.48\textwidth]{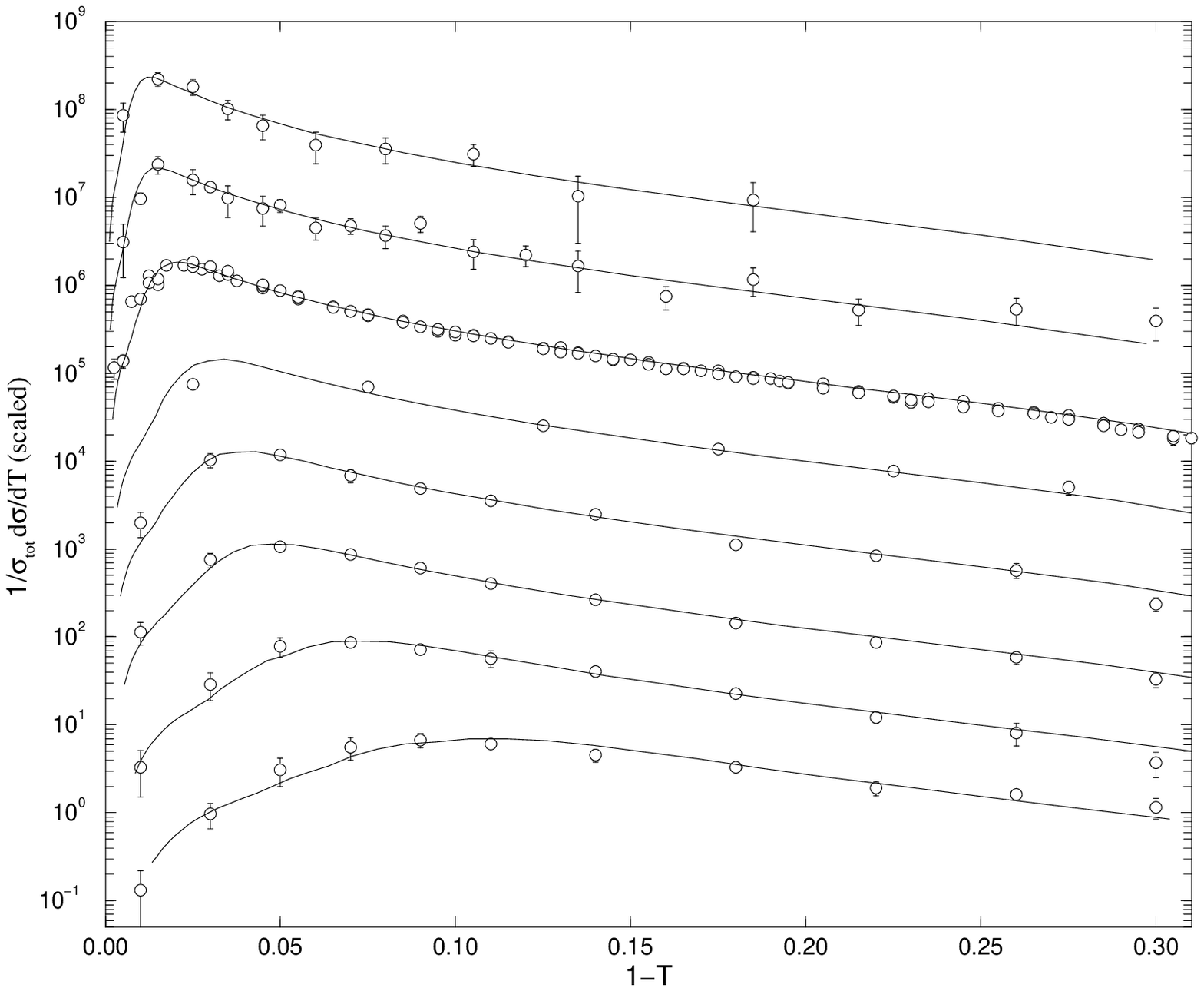}
  \hfill
  \includegraphics[width=0.48\textwidth,height=0.48\textwidth,angle=90]{gr_thrust_kcorr_mz.eps}
  \caption{Left: the thrust distribution using the NLL+NLO perturbative
    distribution with the shape function eq.~(\ref{eq:thrustshape}).
    Shown for $Q=14,22,35,44,55,91,133$ and $161$~GeV. Figure taken
    from~\cite{Korchemsky:1998ev}. Right: the thrust distribution at
    the perturbative (DGE) level, with a non-perturbative shift, and
    with a shape function. Figure taken from~\cite{GRmass}.}
  \label{fig:thrustShape}
\end{figure}

There is considerable freedom in one's choice of the form for the
shape function.  Accordingly, success in fitting a given event-shape
distribution for any single value of $Q$ cannot, alone, be considered
strong evidence in favour of the shape-function picture. The true test
of the approach comes by determining the shape function at one value
of $Q$ and then establishing whether it applies for all values of $Q$.
This is illustrated for the $\ee$ thrust in fig.~\ref{fig:thrustShape}
(left), where the following shape-function has been used
\cite{Korchemsky:1998ev,korch3},
\begin{equation}
  \label{eq:thrustshape}
  f(x) = \frac{2 \left (x/\Lambda\right)^{a-1}}{\Lambda \Gamma \left
      (\frac{a}{2} \right )} \exp\left(-\frac{x^2}{\Lambda^2}\right) \,,
\end{equation}
with two free parameters, $a$ and $\Lambda$ (it has been argued
that the behaviour of this function for small $x$ and large $x$ is
similar to that expected from a simple physical model for the dynamics
at the origin of the shape function~\cite{korch5}).  One notes the
remarkable agreement, far into the two-jet region, for the whole range
of $Q$ values.

This is a clear success for the shape-function approach as compared to
the simple shift discussed above. However the drawback of
shape-functions is that a priori one loses all predictivity (save for
the first moment) when going from one observable to another. This
predictivity was one of the main strengths of the shift approach.
Considerable work has therefore been devoted to understanding both the
general properties of shape functions and the relations between shape
functions of different observables.

One line of investigation has been to relate the shape functions for
the thrust and heavy-jet mass ($\rho_H$) --- this can be done with the
help of the observation~\cite{CTTWlong} that in the two-jet limit the
thrust is just the sum of the light and heavy-jet (squared normalised)
masses. In~\cite{korch4} Korchemsky and Tafat introduced a shape
function ${\tilde f}(x_1,x_2)$ describing the distribution of
non-perturbative radiation in each of the two hemispheres. The shape
function for the thrust is then obtained by
\begin{equation}
  {\tilde f}_T(x) = \int dx_1 dx_2 \tilde f(x_1, x_2) \delta(x - x_1 -
  x_2).
\end{equation}
Instead for the heavy-jet mass one calculates the full (NP) double
differential distribution for the two jet masses and from that derives
the heavy-jet mass distribution (this is not quite equivalent to a
simple shape-function for $\rho_H$). This approach gave a good fit to
the thrust and heavy-jet mass data, on the condition that there was
significant correlation between the non-perturbative contribution to
the two hemispheres.\footnote{The same shape-function parameters were
  also applied to the $C$-parameter (with an overall rescaling to
  ensure the correct first moment), however the data were less well
  described than for the thrust. This is perhaps not surprising: owing
  to different sensitivities to large-angle radiation, the thrust and
  $C$-parameter shape functions are expected to have a different
  structure of higher moments~\cite{GardMan}.} %
It is difficult to interpret this result however, because as has
already been mentioned, the default measurement scheme for the jet
masses suffers from extra (universality-breaking) $\Lambda/Q$
corrections, and it is possible that in~\cite{korch4} these are being
mimicked by inter-hemisphere correlations.

The thrust and heavy-jet mass have been investigated in great detail
also by Gardi and Rathsman~\cite{GRthrust,GRmass}. In analogy with the
approach for the mean thrust~\cite{GardGrun}, they attempt not only to
account for NLL+NLO and power correction contributions, but also for
all (perturbative) enhanced effects associated with the running of the
coupling, within a formalism that they refer to as Dressed Gluon
Exponentiation (DGE). Among the interesting theoretical results (see
also~\cite{Gard}) is that the renormalon analysis suggests that the
second moment of the shape function should be precisely that expected
from a pure shift (the same is true for the $C$-parameter
\cite{GardMan}). This observation seems to be consistent with fits to
the data,\footnote{Though apparently only in the decay (hadron-mass)
  scheme~\cite{GRmass}. It should be kept in mind also that the
  properties of the higher moments may depend on the exact
  prescription for the perturbative calculation.} and would explain
why the shift approximation works so well, at least for the relevant
observables.  The right-hand plot of fig.~\ref{fig:thrustShape} shows
the result of their analysis for the thrust, with both a shift and a
shape-function.  While the shape-function is clearly needed in the
immediate vicinity of the peak and below, it is striking to see, above
the peak, how well it is approximated by the shift.

One of the other interesting results in~\cite{GRmass} relates to the
analysis of the heavy-jet mass. In contrast to~\cite{korch4}, a common
hadron-mass scheme (decay-scheme) is used for the thrust and the
heavy-jet mass, and there are no inter-hemisphere correlations in
their shape function. Using the same parameters for the thrust and
$\rho_H$ leads to reasonable fits, suggesting that the true
inter-hemisphere correlations may actually be small. (A fit for
$\rho_H$ alone does however prefer a lower value of $\as$ than for the
thrust, as is seen also with the usual shifted NLL+NLO approach
\cite{Abdallah:2002xz}).

A puzzle that emerges in the work of~\cite{GRthrust,GRmass} is that
all the fits lead to quite small values of the coupling, $\alpha_s
\simeq 0.109$ with a theoretical error of about $\pm 0.005$. A
similarly low result was obtained in the analogous analysis of the
mean thrust~\cite{GardGrun}. In contrast a study of $b$-fragmentation
using dressed gluon exponentiation finds more standard values,
$\as\simeq 0.118$~\cite{Cacciari:2002xb}.

Let us close our discussion of shape functions with a recent result by
Berger \& Sterman~\cite{BergSterm}. Given that there are no simple
relations between the shape functions for most common event shapes,
they instead consider a new class of event shapes~\cite{BKScorrel}
whose definition involves a parameter $a$ which allows the nature of
the event shape to be continuously varied from a broadening ($a=1$) to
a thrust-type observable ($a=0$) and beyond.  Their observation is
that for all variants of the observable with $a<1$, the logarithm of
any moment of the shape functions scales as $1/(1-a)$. In practice
this scaling is found to hold in Monte Carlo studies of this class of
observables --- it would therefore be interesting to have data for
these observables with which to make a comparison.

\section{Outlook}
\label{sec:outlook}

Conceptually, event shapes are rather simple observables, yet they are
sensitive to a range of characteristics of QCD radiation. This
combination of theoretical simplicity and experimental sensitivity is
perhaps one of the main reasons why event shapes have found so many
applications. As we have seen, they provide vital inputs in studies of
the ingredients of the QCD Lagrangian, such as the coupling and the
colour factors; they play an important role in the tuning and testing
of event generators; and they are at the heart of recent
investigations into analytical approaches to understanding the
dynamics of hadronisation.

There remain several directions in which progress may be expected (or
at the very least hoped for) in coming years. All the experimental
comparisons discussed here have been for event-shapes that vanish in
the $2$-jet limit, yet there have been significant theoretical
developments in recent years for observables that vanish also in the
$3$-jet limit, examples being the $D$-parameter and thrust minor
\cite{BDMZtmin,BDMZdpar,KOUTDIS,KOUTDY,AZIMDIS,NGOneJet,
  Mercutio,MENLO,EERAD2,NLOJET}. Studies with jet rates
\cite{Heister:2002tq} suggest that going to such `three-jet'
observables could reduce theoretical errors, though as discussed in
\cite{Bethke:2002rv} this needs to be investigated more
systematically. Additionally three-jet observables will allow much
more stringent tests of analytical hadronisation approaches since for
the first time they introduce sensitivity to predictions about
hadronisation from a gluon~\cite{BDMZtmin,BDMZdpar}, not just from a
quark.

One of the frontiers of event shape studies is their extension to new
processes. While $\ee$ is the traditional domain for event-shape
studies, we have seen that studies in DIS are in many respects
competitive with $\ee$ results. Additionally they provide important
input on the question of universality of analytical hadronisation
models, ruling out for example significant modifications of the
hadronisation picture stemming from interactions with the proton
remnant. Though so far the question has received only limited study,
one interesting application of DIS studies would be in the use of
event shapes for obtaining information on parton distribution
functions, as suggested in~\cite{KOUTDIS}.

Currently an option that is attracting
growing experimental~\cite{Bertram:sv} and theoretical
\cite{KOUTDY,NLOJETHH,Caesar} interest is the development of event
shapes in hadron-hadron collisions. Though much work remains to be
done on this subject, it seems that the sensitivity to parton
distributions will be somewhat stronger than in most DIS observables.
Furthermore one expects a rich hadronisation structure associated both
with the four-jet structure of hadronic dijet events and with the
properties of the underlying event. It is to be hoped that the
availability of automated resummation methods~\cite{Caesar} will make
the theoretical aspects of the study of new observables and processes
more straightforward than has been the case in the past.

Another frontier on which progress is expected in near future is with
respect to the accuracy of theoretical predictions. NNLO perturbative
calculations are making rapid progress (for a recent review see
\cite{GehrmannReview}) with the main outstanding problem being that of
a full subtraction procedure for combining zero, one and two-loop
contributions and its implementation in a Monte Carlo integrator.
Progress on NNLL resummations is also being made, albeit so far only in
inclusive cases~\cite{BCFG}. Though technically more involved, an
extension to event shape resummations is certainly conceivable in the
near future. Aside from the expected gains in accuracy that are the
main motivation for these calculations, it will also be interesting to
compare the exact higher-order calculations with predictions from
approaches such as~\cite{GRthrust,GardGrun,Maxwell} which aim to
identify the physically dominant higher-order contributions and in
some cases~\cite{Maxwell,Abdallah:2002xz} claim to significantly
reduce the theoretical uncertainties on $\as$.

Less predictable is what development can be expected on the subject of
hadronisation corrections. As we have seen, much has been learnt from
event-shape studies. Yet a variety of open questions remain. At a
practical level it would for example be of interest to see an
extension of the shape-function approach to a wider range of
variables, including the quite subtle case of the broadenings and
also to DIS event shapes. The wealth of DIS data at low $Q$, where
shape functions are most important, makes the latter especially
interesting.

There remain also important issues that are poorly understood even at
the conceptual level. For example predictions are currently limited to
the domain in which the event-shape is close to vanishing (e.g.\ $1-T$
in the $2$-jet limit).  It is tempting to suggest that problems that
persist for a couple of `recalcitrant' observables (specifically,
$\rho_H$, $B_W$) may actually be due to large differences between the
power correction in the three-jet region (a significant part of the
fit-range) and that in the well-understood two-jet limit (see e.g.\ 
fig.~3 of~\cite{Kluth:2003uq}).  But techniques that would allow the
calculation of the power correction to, say, the thrust or the
broadening in the $3$-jet 
limit do not yet exist.  Another (partially related) issue is
that of anomalous dimensions --- these have so far only been
calculated for $\Lambda/Q$ effects that are associated with hadron
masses~\cite{Salwick}. Yet, as has been pointed out also in
\cite{Magnea:2002xt}, they are bound to be present for all classes of
power suppressed contributions. One clear physical origin for them is
that soft-gluons (or hadrons) with transverse momenta of order
$\Lambda$ are radiated not just from the $q\bar q$ system but also
from all soft perturbative gluons with transverse momenta between
$\Lambda$ and $Q$.  Accounting for radiation from the latter will
clearly lead to enhancements of power corrections by terms $(\as \ln
Q/\Lambda)^n$. And a final point, not to be forgotten, is that most of
the analytical hadronisation approaches are essentially perturbative
methods `in disguise'. An understanding of how they really relate to
\emph{hadron}isation is very much lacking.

Let us close this review by remarking that one of the characteristics
of event shape studies is that progress is made not merely through a
better understanding of existing observables and data, but also
through critical experimental studies of new observables with
(supposedly!) better
theoretical properties. The dialogue between experimenters and theorists
is crucial in this respect. We look forward to its being as fruitful
in coming years as it has been up to now.

\section*{Acknowledgements}

It goes without saying that we are grateful to our collaborators and
numerous other colleagues, including members of the LEP QCD working
group, for the many stimulating discussions that we have had about
event shapes.  For discussions on issues that arose specifically while
writing this review, we thank in particular G.~Heinrich, F.~Krauss,
C.~J.~Maxwell, G.~Rodrigo, M.~H.~Seymour, H.~Stenzel and
G.~Zanderighi.  Furthermore we are indebted to A.~Banfi,
Yu.~L.~Dokshitzer and G.~Zanderighi for their careful reading of and
helpful comments on the manuscript.

One of us (GPS) is grateful also to the CERN TH division for
hospitality during the course of this work.


\end{document}